\documentclass[iop,apj,twocolappendix,numberedappendix,twocolumn,usenames,dvipsnames]{aastex63}

\newbox\grsign \setbox\grsign=\hbox{$>$}
\newdimen\grdimen \grdimen=\ht\grsign
\newbox\laxbox \newbox\gaxbox
\setbox\gaxbox=\hbox{\raise.5ex\hbox{$>$}\llap
     {\lower.5ex\hbox{$\sim$}}}\ht1=\grdimen\dp1=0pt
\setbox\laxbox=\hbox{\raise.5ex\hbox{$<$}\llap
     {\lower.5ex\hbox{$\sim$}}}\ht2=\grdimen\dp2=0pt

\newcommand{\rev}{\color{black}}

\usepackage[T1]{fontenc}
\usepackage{amsmath}
\usepackage{amssymb}
\usepackage{ctable} 
\usepackage{graphicx}
\usepackage{natbib}
\usepackage[titletoc,title]{appendix}
\usepackage{ulem}

\usepackage{mathtools}
\usepackage{lmodern}
\usepackage{csquotes}
\usepackage{xspace}

\begin{document}

\title{Particles in Relativistic MHD Jets. I. Role of Jet Dynamics in Particle Acceleration}
\shorttitle{Jet Particle Acceleration}
\shortauthors{Dubey et al.}

\author[0000-0002-8506-9781]{Ravi Pratap Dubey}
\altaffiliation{Fellow of the International Max Planck Research School for Astronomy \& Cosmic Physics at the University of Heidelberg}
\affiliation{Max Planck Institute for Astronomy, K{\"o}nigstuhl 17, D-69117 Heidelberg, Germany}
\email{dubey@mpia.de,  fendt@mpia.de, bvaidya@iiti.ac.in}

\author[0000-0002-3528-7625]{Christian Fendt}
\affiliation{Max Planck Institute for Astronomy, K{\"o}nigstuhl 17, D-69117 Heidelberg, Germany}

\author[0000-0001-5424-0059]{Bhargav Vaidya}
\affiliation{Indian Institute of Technology Indore, Khandwa Road, Simrol, Indore 453552, India}
\begin{abstract}
Relativistic jets from (supermassive) black holes are typically observed in non-thermal emission, caused by highly-relativistic electrons. Here, we study the interrelation between three-dimensional (special) relativistic magnetohydrodynamics, and particle acceleration in these jets. We inject Lagrangian particles into the jet that are accelerated through diffusive shock acceleration and radiate energy via synchrotron and inverse Compton processes. We investigate the impact of different injection nozzles on the jet dynamics, propagation, and the spectral energy distribution of relativistic particles. We consider three different injection nozzles -- injecting steady, variable and precessing jets. These jets evolve with substantially different dynamics, driving different levels of turbulence and shock structures. The steady jet shows a strong, stationary shock feature, resulting from a head-on collision with an inner back-flow along the jet axis - a jet inside a jet. This shock represents a site for highly-efficient particle acceleration for electrons upto a few tens of TeV and should be visible in emission as a jet knot. Overall, we find that the total number of shocks is more essential for particle acceleration than the strength of the shocks. The precessing jet is most efficient in accelerating electrons to high energies reaching even few hundred TeVs, with power-law index ranging from 2.3 to 3.1. We compare different outflow components, such as jet and the entrained material concerning particle acceleration. For the precessing nozzle, particle acceleration in the entrained material is as efficient as in the jet stream. This is due to the higher level of turbulence induced by the precession motion.
\end{abstract}

\keywords{
   Active galactic nuclei(16) --
   Relativistic jets(1390) -- 
   Magnetohydrodynamics(1964) --
   High energy astrophysics(739) --
   Particle astrophysics(96) --
   Shocks(2086)
 }

\section{Introduction}
\label{sec:intro}
Astrophysical jets are often observed as relativistic, well-collimated outflows of plasma and fields originating in a region 
near a compact object accreting matter. 
These jets are present in a number of astrophysical systems on various scales including Active Galactic Nuclei (AGNs), high mass X-ray 
binaries or micro-quasars, gamma-ray bursts (GRBs) and young stellar objects (YSOs). 
Jets originate from a deep gravitational potential well and from sources hosting strong magnetic fields and an accretion disk \citep{hawley2015}. 

AGN jets travel long distances, reaching hundreds of kpc from the supermassive black hole. 
The parent galaxies of these sources (called radio galaxies), can be categorized into two classes \citep{FR1974} depending on their 
radio jet kinetic  power and morphology. 

Although the processes responsible for formation and launching of these jets are not yet completely understood, the 
power released from the rotational energy of the black hole \citep{BZ1977} or the accretion flow \citep{BP1982} are
thought to be the sources of jet energy. 
This power is ultimately transferred from the gravitational energy to the jet kinetic energy via the magnetic field.
The magnetic field can be deduced from observations of power-law spectra, polarization 
spectral distribution and polarisation. 
The magnitude of the magnetic field is typically obtained from the cutoff in power-law spectra, whereas the direction is determined via polarisation of these systems,
\citep{meisenheimer1997, carilli1999, heavens1987, brunetti2003} 
that is due to non-thermal particles that undergo synchrotron and inverse Compton losses.

AGN jets are typically super-(magneto)sonic, and hence produce strong shocks leading to a turbulent magnetic field and velocity component. 
This makes these sources an ideal site for accelerating particles to high energies \citep{hargrave1974}. 
Strong shocks, through diffusive shock acceleration, play a vital role in accelerating particles \citep{krymskii1978, blandford1978, bell1978a, bell1978b}. 
Additionally, various other mechanisms of particle acceleration are also at play in these sources. 
These include second-order Fermi acceleration \citep{fermi1949, kundu2021} particularly in the radio lobes, magnetic reconnection in highly magnetized 
jet regions \citep{giannios2009} and shear acceleration \citep{rieger2007, sironi2021}.

The dynamics of the jet seems to plays an essential role in the emission and subsequently, in studying the shock structure and particle acceleration.
In particular, jets that undergo a time-dependent injection of material (and magnetic field), such as intermittent jets and precessing jets,
are able to drive a high excess of turbulence or instabilities in the jet flow. 
This leads to more shocks, and hence, more sites of particle acceleration, and also to an increased efficiency 
of particle acceleration.
\citep{yates2018, giri2022}. 

The AGN jet sources are also thought to be one of the sites for production of ultrahigh energy cosmic rays \citep{PAC2017, aab2018, matthews2018}. 
Although several studies have pointed out the ineffectiveness of highly relativistic shocks to accelerate particles to such high energies \citep{reville2014, bell2018}, 
other investigations have shown that non-relativistic or mildly relativistic shocks can accelerate particles to high energies 
\citep{kirk1999, marcowith2016, Matthews2019, marcowith2020, araudo2021, ortuno-macias2022}.  
Such shocks are capable of accelerating both protons and electrons. 
In particular, existence of high energy protons is crucial for understanding the origin of neutrino emission associated with blazar jets \citep{mannheim1993}. 
Even at large scales, these accelerated particles play a crucial role in governing the heating due to feedback. 
Simulations of jet inflated bubbles in presence
of cosmic ray transport have shown the dominant role of cosmic ray mediated heating at large cluster scales \citep{ehlert2018}.  

As the propagation of astrophysical jets is highly dynamical due to the various non-linear physical processes at work, numerical simulations play 
an essential role for studying jets. 
A number of numerical simulations have been performed to study various aspects of jet propagation
(for relativistic jets see e.g. \citealt{leismann2005,mizuno2007, Keppens2008, rossi2008, meliani2009, Mignone2010}). 
In particular, shock diagnostics and  particle acceleration in the jet are studied in \citet{Mukherjee2021} for a straight jet with small 
velocity perturbations to induce turbulence through three-dimensional (3D) relativistic magnetohydrodynamical (RMHD) simulations.

In our paper, we present 3D (special) RMHD simulations of jets injected into {\rev a uniform} ambient medium.
We inject Lagrangian {\rev macro-particles into the propagating jet that are advected along with the flow, each of them
representing an ensemble of electrons.
We then look how these particles are accelerated along the jet flow 
by internal and external shocks \citep{vaidya2018}}.
We study different kinds of time-dependent injection nozzles and investigate how injection from these nozzles affects the dynamics of the 
jets - its turbulence, shock structure, symmetry - and, as a result, how particle acceleration varies for these nozzles.
More specifically, we use three different nozzles -- injecting 
(i) a straight jet, 
(ii) a precessing jet and 
(iii) an intermittent jet into the ambient medium.  

The different nozzles we apply are thought to be generated by different jet launching conditions at the jet origin.
Jet launching itself - the transition from accretion to ejection - is not further considered in our simulations.
We may distinguish between jets being launched from resistive accretion disks, as modeled in the non-relativistic case in axisymmetry
(see e.g. \citealt{keppens2002, zanni2007, sheikhnezami2012, stepanovs2016}).
Fully 3D simulations e.g. for binary systems in a Roche potential \citep{sheikhnezami2015} indicate on precession effects on the accretion disk
and thus the jet launched from that disk.

Jet launching from the black hole ergosphere (the BZ process) has been numerically modeled since two decades, mostly applying the evolution
from an initial torus surrounding the black hole \citep{mckinney2004, tchekhovskoy2011}.
More recent work applying resistive GR-MHD has been able to consider in addition the evolution of disk winds and jets from thin
accretion disks and comparing them to the central spine jet \citep{qian2018,vourellis2019, dihingia2021}, even including a disk dynamo \citep{vourellis2021}.
We note that the kinematics of the BZ jets obtained in these simulations is biased by the floor model
for density and pressure that needs to be considered.

The observation of jet knots 
hints to a time-dependent mechanism behind their origin, probably a time-dependent injection (nozzle).
Some recent observations indeed provide indication for precessing AGN jets (see e.g. \citealt{britzen2018,vonFellenberg2023}),
and their connection to binary black holes (see e.g. \citealt{kharb2017,krause2019}).

Our paper is structured as follows. 
In section~\ref{sec:model} we detail our model approach and the numerical methods.
In section~\ref{sec:res_dynamics} we discuss the dynamical evolution of the different jets, their
energetics and turbulence level.
In section~\ref{sec:res_particles} we finally present energy spectra of particles moving with the jet and being 
accelerated by shocks arising in the turbulent jet motion.
We conclude with a summary in section~\ref{sec:summary}.

\section{Model Setup}
\label{sec:model}
We apply the PLUTO code \citep{mignone2007pluto} to solve the set of (special) relativistic, magnetohydrodynamical (RMHD) fluid equations 
\begin{equation}
    \frac{\partial}{\partial t} \begin{pmatrix}
        D \\ \textit{\textbf{m}}  \\ E_t \\ \textit{\textbf{B}}
    \end{pmatrix} 
    + \nabla \cdot \begin{pmatrix}
        D\textit{\textbf{v}} \\ w_t \gamma_{\rm f}^2\textit{\textbf{vv}} - \textit{\textbf{bb}} +  p_t \bar{{\textbf{{I}}}}   \\ \textit{\textbf{m}} \\ \textit{\textbf{vB}} - \textit{\textbf{Bv}}
    \end{pmatrix}^T = 0
\end{equation}
on a three-dimensional (3D), uniform Cartesian grid. 
Here, $D$ is the laboratory density, 
$\textit{\textbf{m}}$ is the momentum, 
$\textit{\textbf{B}}$ is the magnetic field in the lab frame, 
$E_t$ is the total energy density, 
$\textit{\textbf{v}}$ is the velocity,
$\gamma_{\rm f}$ is the Lorentz factor of the fluid, and 
$\bar{\textbf{I}}$ is the diagonal tensor. 
The magnetic field 
can be represented by the 4-vector 
\begin{equation}
b^{\mu} \equiv
\left[b^0, b^i \right] = 
\left[\gamma_{\rm f} \textit{\textbf{v}} \cdot \textit{\textbf{B}}, 
     \ \gamma_{\rm f} \left(\frac{B^i}{\gamma_{\rm f}^2} + v^i(\textit{\textbf{v}} \cdot \textit{\textbf{B}})\right) \right].
\end{equation}
with the magnetic energy density
\begin{equation}
     b^2 = b^\mu b_\mu = \frac{B^2}{\gamma_{\rm f}^2} + (\textit{\textbf{v}} \cdot \textit{\textbf{B}})^2, 
\end{equation}
{\rev the momentum vector $m^i = w_t \gamma_{\rm{f}} v^i-b^0b^i $},
the relativistic total enthalpy $w_t = \rho h + b^2$, and the total pressure of the fluid $p_t = p + (b^2/2)$ where $p$ is the gas pressure \citep{delzanna2003, mignone2007taub}.
Further, the specific enthalpy $h$ is related to the internal energy $\varepsilon$ of the gas as 
\begin{equation}
     h = 1 + \varepsilon + \frac{p}{\rho},
\end{equation}
where $\rho = D/\gamma_{\rm f}$ is the density in the fluid frame and $p$ is the gas pressure. 

In order to close the aforementioned equations, an equation of state relating $h$ with $\rho$ is further needed 
(which we discuss in  Section~\ref{sec:eos}). 
With this, the total energy density in the lab frame is given as 
\begin{equation}
\label{eq: E_total}
    E_t = Dh\gamma - p + \frac{|\textit{\textbf{B}}|^2}{2} + \frac{|\textit{\textbf{v}}|^2 |\textit{\textbf{B}}|^2 - (\textit{\textbf{v}} \cdot \textit{\textbf{B}})^2}{2}
\end{equation}
\citep{mignone2007taub}, where the first two terms represent the kinetic, thermal and the rest mass energy density while the last two terms represent the electromagnetic energy density. 
Specifically, the third term represents the magnetic energy density, whereas the last term represents the electric energy density as a result of an electric field ${\bf \mathcal{E}} = - $\textit{\textbf{v}$\times$\textbf{B} } \citep{Keppens2008}, altogether considering the contribution by Poynting flux.

To study the particle acceleration, we use the Lagrangian particle module of the PLUTO code \citep{vaidya2018}, 
which uses passive Lagrangian particles moving in an Eulerian grid to model the non-thermal spectral signatures from relativistic jets. 
Essentially, we consider particle acceleration due to diffusive shock acceleration (DSA; or first-order Fermi acceleration) following the sub-grid approach 
developed in \cite{vaidya2018, Mukherjee2021}.
The Lagrangian macro-particle, in this approach, represents an ensemble of particles (electrons) following the fluid, initially 
distributed as following a power-law energy spectrum. 
This initial spectrum subsequently evolves for each macro-particle in time and momentum space, taking into account acceleration due to DSA and cooling as a result of adiabatic expansion, 
synchrotron radiation processes, and inverse Compton (IC) scattering of particle in a background of cosmic microwave background (CMB) photons.
For this purpose, the particle code solves the relativistic transport equation for cosmic rays in scattering medium \citep{webb1989} 
{\rev which we discuss in Section~\ref{sec:radiation}}. 

\subsection{Equation of State}
\label{sec:eos}
In order to solve the RMHD equations mentioned above, we need a proper closure provided by an additional equation, 
that is the equation of state (e.o.s.), relating two thermodynamic parameters (e.g.~density and internal energy). 

For a perfect gas, this relationship is derived using the relativistic theory \citep{Synge1957}. 
However, applying this general relationship in a numerical code is too time-consuming and for saving computational resources, 
a constant-$\Gamma$ e.o.s. is often used.
We apply an approximate e.o.s. which follows a simple analytical form and is fast and suitable to be adopted for 
numerical studies,
\begin{equation}
\label{eq:h_tm}
    h = \frac{5}{2} \Theta + \sqrt{\frac{9}{4}\Theta^2 + 1},
\end{equation}
\citep{mathews1971,Mignone2005} where $\Theta = p/\rho$ is the temperature.
This e.o.s., widely denoted as Taub-Mathews (TM) e.o.s., differs from the theoretical Synge e.o.s. only by a few percent, 
and provides thermo-dynamical variables that are very similar to the constant-$\Gamma$ e.o.s. in the limit of a cold 
gas ($\Gamma =5/3$, $\Theta \to 0$) and a hot gas ($\Gamma =4/3$, $\Theta \to \infty$). 
For intermediate temperatures, the respective values vary smoothly between the two limiting cases. 
For the TM e.o.s., the adiabatic index is  
$\Gamma = (h-1)/(h-1-\Theta)$,
whereas the sound speed relative to the fluid $c_{\rm s}$ is defined as
\begin{equation}
\label{eq:cs}
    c_{\rm s}^2 = \frac{\Theta}{3h}\frac{5h-8\Theta}{h-\Theta}.
\end{equation}
Applying $c_{\rm s}$ we define the (ordinary) relativistic sonic Mach number in the lab frame $M_{\rm s} = v/c_{\rm s}$ and
the proper sonic Mach number 
\begin{equation}
\label{eq:mach}
   \mathcal{M}_{\rm s} \equiv \frac{u}{u_{\rm s}} = \frac{\gamma_{\rm f} v}{\gamma_{\rm fs} c_{\rm s}} = \frac{\gamma_{\rm f}}{\gamma_{\rm fs}} M_s
\end{equation}
with the proper speed of the fluid $u$ and the proper sound speed  $u_{\rm s}$ relative to the fluid
(and similarly for the Lorentz factors $\gamma_{\rm f}$ and $\gamma_{\rm fs}$, respectively).
Here we follow the arguments by \citet{koenigl1980} and others stating that the proper sonic Mach number $\mathcal{M}_{\rm s}$
is the relativistic generalization of the Newtonian Mach number. 
Similarly, the Alfv\'en velocity $c_{\rm a}$, the magnetisation $\sigma$ and the plasma $\beta$ are defined as 
\begin{equation}
    c_{\rm a}^2 = \frac{b^2}{\rho h + b^2} \ , \ \ \sigma = \frac{b^2}{\rho c^2}, \ \ \beta = \frac{p}{b^2},
\end{equation}
and similarly for the related Alfv\'en Mach numbers $M_{\rm a}$ and $\mathcal{M}_{\rm a}$ 
\citep{mizuno2014}. 

As {\em extractable} energy $E_{\rm x}$ of the fluid in the lab frame, we define
the total energy $E_{\rm t}$ (Equation~\ref{eq: E_total}) subtracted by the rest mass energy density $Dc^2$.
Substituting the specific enthalpy $h$ for the TM e.o.s we obtain
\begin{eqnarray}
\label{eq:E_x}
    E_{\rm x} & = & \gamma(\gamma-1)\rho c^2 + \left[(h-1)\gamma^2 - \Theta \right] \rho  \nonumber  \\
        & + & \frac{B^2}{2} + \frac{|\textit{\textbf{v}}|^2 |\textit{\textbf{B}}|^2 - (\textit{\textbf{v}} \cdot \textit{\textbf{B}})^2}{2}
\end{eqnarray}
where the first and second terms on the right hand side represent 
the kinetic energy density $E_{\rm kin}$ and 
the thermal energy density $E_{\rm th}$, respectively,
and the last two terms represent the electromagnetic energy density $E_{\rm em}$.
The extractable power of the jet $P_{\rm x}$ at time $t$ is calculated by integrating the extractable energy density $E_{\rm x}$ over the 
volume $V$ and dividing by the time $t$.
Thus, $P_{\rm x} = (\int E_{\rm x}dV)/t$ = $\overline{E}_{\rm x} V /t$, where $\overline{E}_{\rm x} = \int (E_{\rm x}dV)/V$ is the (volume-weighted) mean extractable energy
density over the outflow volume $V$ (considering cells, where $v>0)$.

We have thoroughly tested both ideal and TM (implemented in PLUTO as Taub e.o.s.) e.o.s. in our simulations in order to compare 
their effects on the flow dynamics and internal structure.
We came to the conclusion that the TM e.o.s. is a much better approach when treating (relativistic) gases of different
temperature, such as the low-density, hot gas in the jet, and the high-density, cool gas of the ambient medium.
The TM approach is particularly interesting when it comes to the dynamics of the jet cocoon or back-flow material consisting of 
shocked and entrained gas, thus a mixture of jet gas and ambient gas. 

\subsection{Numerical Specifics}
\label{sec:num}
Applying a 3D Cartesian grid, we model magnetized, rotating, one-sided relativistic jets injected from different injection
nozzles.

In order to be able to study the turbulence and shock properties properly, it is necessary to resolve small-scale motions as
good as possible.
We have thus performed a resolution study by modeling and comparing the evolution and structure of the jet at different grid resolution
(see Appendix~\ref{appendix:resolution}) {\rev applying the same input profiles and parameters values as specified
for the steady, straight jet setup in the next section}.
For the resolution study we have compared simulations with a resolution of $5$, $10$, $15$, $20$, $25$, and $30$ cells 
per jet radius $r_j$. 
We find that while the jet dynamical variables as well as particle energy spectrum are sufficiently converged for a resolution of 25 grid cells per jet radius.  

We use divergence cleaning to implement the solenoidal condition for the magnetic field 
$\nabla \cdot \textit{\textbf{B}} = 0$.
We adopt a multidimensional shock flattening algorithm, a second order Runge-Kutta time-stepping, a Courant-Friedrichs-Levy number 
of 0.25, and a HLL Riemann solver with linear reconstruction that is second order accurate in space. 
 
While the MHD simulations are basically scale-free and could be applied to a variety of sources of different size
and energy output, the treatment of radiation {\rev as well as of particles} requires a proper astrophysical scaling, 
thus adapted to individual sources of interest.
Radiation effects, such as particle acceleration and cooling as well as determining emissivities and intensities, depend on actual density, 
pressure (temperature) or magnetic field strength.
For the present paper, since we are mostly interested in the dynamics and the particle acceleration, 
we work in dimensionless code units for most of the time (except where we explicitly mention the physical units). 
In order to convert from code units to astrophysical units, a suitable scaling factor must be applied.
However, the synchrotron cooling modeled in our simulations depends on the physical value of the magnetic field.
In order to deal with that, we have chosen one set of such normalization units in our simulations (see Table~\ref{tbl:input_normalization}).
The different variables evolved in our code are normalized based on the choice of three basic unit parameter,
the unit length $l_0 = r_j$, the unit speed $v_0=c$ and the unit density $\rho_0$. 
Other variables or the normalized accordingly as 
$t_0 = l_0/v_0$, 
$B_0 =  v_0 \sqrt{4 \pi \rho_0}$,
$p_0 = \rho_0 v_0^2$,
$T_0 = \mu m_u v_0^2/2 k_B$,
where $t_0$, $B_0$, $p_0$ and $T_0$ represent the unit time, the unit magnetic field, the unit pressure and temperature, 
respectively,
while $m_u$, $\mu$ and $k_B$ denote the atomic mass unit, the mean molecular weight and the Boltzmann constant, 
respectively\footnote{Note that the factor $\sqrt{4 \pi}$ is included in the definition of the magnetic field in 
PLUTO {\rev intrinsically. Hence, the factor of $\sqrt{4 \pi}$ is explicitly factored in the definition of $B_0$ as well to compensate.}}. 
{\rev The unit for the power is then defined as $P_0 = l_0^3 \rho_0 v_0^2 / t_0$.}

\begin{deluxetable}{cc}

\tabletypesize{\scriptsize}
\tablewidth{0pt}
\tablecaption{Normalization Units \label{tbl:input_normalization}}
\tablehead{
\colhead{Parameter} & \colhead{Conversion Factors}
}

\startdata
$l_0$     &    $0.5 \ {\rm pc}$ \\
$v_0$  &  $2.998 \times 10^{10} \ \rm{cm\,s}^{-1}$ \\
$\rho_0$  & $1.66 \times 10^{-26} \ \rm{g\,cm}^{-3}$ \\ 
\noalign{\smallskip}\hline \noalign{\smallskip}
$t_0$  & $1.63 \ \rm{yr}$ \\ 
$B_0$  & $13.69 \ \rm{mG}$  \\ 
$p_0$  & $1.49 \times 10^{-5} \ \rm{dyne\,cm}^{-3}$ \\
$T_0$  & $5.41 \times 10^{12} \ \rm{K}$  \\
$P_0$  & $1.07 \times 10^{42}\ \rm{ergs\,s^{-1}}$  \\
\enddata    
\tablecomments{Conversion factors from code units to physical scales.
Shown for the three basic parameters: length $l_0$, speed $v_0$ and density $\rho_0$ along with 
the derived normalization factors for time $t_0$, magnetic field $B_0$, pressure $p_0$, temperature $T_0$, and power $P_0$.}

\end{deluxetable}

\subsubsection{Initial Conditions}
\label{sec:initial_condition}
We employ a constant density profile across the domain except within a cylindrical region inside the domain, 
with a radius $r_{\rm j} =1$, a height $z_{\rm j} =1$, centered at $x=y=z=0$. 
This is the \textit{injection nozzle}.
Initially, the \textit{ambient medium} is defined as the region outside the injection nozzle
with a density $\rho_{\rm a} = 1000$, and being at rest, $\textit{\textbf{v}}_{\rm a}=0$. 
The initial gas pressure $p =0.1$ is constant throughout the domain (including the injection nozzle). 

The magnetic field in the ambient medium is purely vertical,   $B_{z,\rm a} = 0.176$, thus $B_{r, \rm a} = B_{\phi, \rm a} = 0$.
Here, $B_{r, \rm a}$, $B_{\phi, \rm a}$ and $B_{z, \rm a}$ are the components of the magnetic field in the ambient medium in cylindrical coordinates $r$, $\phi$ and $z$, respectively.

\subsubsection{Boundary Conditions -- Steady Jet Equilibrium}
\label{sec:boundary_conditions}
We investigate three different injection nozzles -- a steady injection, 
and two nozzles with time-dependent injections giving -- a jet with a variable velocity and a precessing jet.
Here, we first describe the boundary conditions for the steady jet injection.
A time variation of these will be applied for the other two nozzles (see next sub-section).

In all our simulations, we inject an under-dense jet with density $\rho_j =1$ from an {\em injection nozzle} of radius $r_{\rm j} =1$ and
height $z_{\rm j} =1$, centered at $x=y=z=0$.
The nozzle is prescribed by a PLUTO user-defined internal boundary condition.

We have implemented outflow boundary condition at all other boundaries of the domain for all setups, applying a zero-gradient condition for all variables.
Through this, we make sure that the material at the boundary can leave the domain freely. 
Note that using such boundary condition may give rise to an inward Lorentz force in case of sub-fast magnetosonic flows, 
resulting in artificial collimation of the jet \citep{porth2010}. 
To avoid this, we restrict the outflow boundaries sufficiently far from the jet stream, not allowing the jet material to leave the domain during the simulation. 

Since our main goal is to investigate how the physics and geometry of the nozzle impacts the jet propagation and subsequent high energy particle 
acceleration,
we find it essential to apply a verified equilibrium solution for the gas and the magnetic field that is injected through the nozzle. 
Any internal instability of the injected gas may lead to an (artificial) modification of the internal equilibrium.

As a proper equilibrium condition for the jet {\rev injection we apply the equilibrium solution derived by \citet{bodo2019} 
inside the jet injection nozzle}. 
For convenience of the reader, we show the corresponding equations for the leading variables (e.g. velocity, magnetic field) along
with plots of the injection profiles in Appendix~\ref{appendix: inj_profiles}. 
Applying these profiles for our simulations, we obtain a stable jet injection with the bulk motion in the selected injection direction
(z-axis) and a rotation around the jet axis.
We refer to this simulation setup with steady injection as {\em std} throughout the text. 

\subsubsection{Boundary Conditions -- Time Variation}
\label{sec:prec_jet}
We now describe the time-dependent nozzles for a straight jet with a time-variable injection and a precessing jet.
For the {\em time-variable injection}, we choose to vary the velocity of the jet in the $z-$direction, applying a sinusoidal variation above a floor velocity,
\begin{equation}
    v'_z (t) = v'_{z,\rm flr} + v'_{0}\cos^2(\omega_{\rm var} t),
\end{equation}
where $v'_{z,\rm flr}$ is a floor velocity (comparable to the other two setups),
$v'_0$ is the amplitude of the variable component of $v'_z$, and
$\omega_{\rm var}$ is the frequency of the variation. 
Here, we set $v'_{z,\rm flr}$ and $v'_0$ at a value of $80$ and $20$ percent of $v_z$, the velocity of the steady jet in $z-$direction, respectively. 
{\rev The maximum velocity (in $z-$direction) of the injected jet in variable nozzle setup is then same as that of the steady nozzle setup, 
corresponding to a Lorentz factor $\gamma_{\rm f} = 10$ at the jet axis.}
Through this we get a straight jet with a velocity varying sinusoidally with a period $\mathcal{P}_{\rm var} = 2\pi/\omega_{\rm var} = 5$ {\rev in the $z-$direction}.
We refer to this simulation setup with variable injection as {\em var} throughout the text.

The setup for the {\em precessing nozzle} is more complex as we need to change that vector components of velocity and magnetic field periodically,
while maintaining the equilibrium solution across the nozzle.
We thus need to
introduce additional terms in the velocity vector of the rotating jet described in previous section. 
We do this by using transformation matrices $\mathbb{T}_x$, which represents a rotation by an angle $\psi$ about the $x-$axis, and $\mathbb{T}_z$, 
which represents a time dependent rotation at an angular velocity $\omega_{\rm prc}$ about the $z-$axis. 
Using these transformation matrices, we get the precessing velocity vector
\begin{equation}
    \begin{pmatrix}
         v'_x \\ v'_y \\ v'_z
    \end{pmatrix}
    =
    \mathbb{T}_z \ \mathbb{T}_x \ 
    \begin{pmatrix}
         v_{x} \\ v_{y} \\ v_{z}
    \end{pmatrix}
\end{equation}
with
\begin{equation}
    \mathbb{T}_x = \begin{pmatrix}
              1 & 0  & 0 \\ 
              0 & \cos{\psi}  & -\sin{\psi} \\ 
              0 & \sin{\psi}  & \cos{\psi}
           \end{pmatrix}
\end{equation}
and
\begin{equation}
     \mathbb{T}_z = \begin{pmatrix}
               \cos{\omega_{\rm prc} t} & -\sin{\omega_{\rm prc} t}  & 0 \\ 
               \sin{\omega_{\rm prc} t} & \cos{\omega_{\rm prc} t}  & 0 \\ 
               0 & 0 & 1
           \end{pmatrix}
\end{equation} 
Here, the un-primed velocity components correspond to the non-precessing jet described in the previous section and $t$ is time. 
{\rev Multiplying this unprimed velocity vector gives us a jet with a velocity vector having the same magnitude as the unprimed velocity vector, but now inclined by
an angle $\psi$ from the $z-$axis. 
Subsequently multiplying this resultant {\em tilted} velocity vector by $\mathbb{T}_z$, we rotate this velocity vector with an angular velocity $\omega_{\rm prc}$ in the $x-y$ plane.}
Hence, we get a rotating jet which is precessing at a precession angle $\psi = 10^o$ about the $z-$axis with a precession period 
$\mathcal{P}_{\rm prc} = 2\pi/\omega_{\rm prc} = 5$. 
We refer to this simulation setup with precessing injection as {\em prc} throughout the text.
{\rev Note that the angular frequency of $\omega_{\rm var}$ corresponds to a sinusoidal variation of velocity in a particular direction i.e the jet axis.
On the other hand $\omega_{\rm prc}$ refers to a precessional motion of the injected velocity  vector.}

We present the input values of selected parameters which we have chosen for jet profiles as mentioned in Appendix~\ref{appendix: inj_profiles}, along with other initial values for some 
jet parameters applied in the jet nozzle at $t=0$, as well as the initial parameters for particle injection (discussed in the next section) in Table~\ref{tbl:input}. 
Note that these values are common to all three injection nozzles as we prescribe the same initial boundary conditions for all the nozzles.

\begin{deluxetable*}{ccccccccccccccccccccc}
\tabletypesize{\scriptsize}
\tablewidth{0pt}
\tablecaption{Input and Initial Parameters \label{tbl:input}}
\tablehead{
\colhead{Parameter} & \colhead{$\rho_j$} & \colhead{$\rho_a$} & \colhead{$p$} & \colhead{$B_{zc}$} & \colhead{$\delta_c$} & \colhead{$\Omega_c$} & \colhead{$\gamma_{c}$} & \colhead{$\overline{\mathcal{M}}_{\rm s}$} & \colhead{$\overline{\mathcal{M}}_{\rm a}$} & \colhead{$\overline{\sigma}$} & \colhead{$\overline{\beta}$} & \colhead{$\overline{\gamma}$} & \colhead{$P_{\rm x}$} & \colhead{$E_{\rm kin}$} & \colhead{$E_{\rm th}$} & \colhead{$E_{\rm em}$} & \colhead{$\alpha$} & \colhead{$\gamma_{\rm min}$} & \colhead{$\gamma_{\rm max}$} & \colhead{$\epsilon$}\\
\colhead{} & \colhead{} & \colhead{} & \colhead{} & \colhead{} & \colhead{} & \colhead{} & \colhead{} & \colhead{} & \colhead{} & \colhead{} & \colhead{} & \colhead{} & \colhead{} & \colhead{$\times 10^6$} & \colhead{$\times 10^6$} & \colhead{$\times 10^6$} & \colhead{} & \colhead{($\log$)} & \colhead{($\log$)} & \colhead{}
}
\colnumbers
\startdata
Value & 1 & 100 & 0.1 & 0.18 & 0.01 & 0.2 & 10 & 24 & 16 & 0.67 & 0.25 & 8.8 & 407 & 3.6 & 1.0 & 1.7 & 6 & 2 & 8 & 0.009 \\ 
\enddata    
\tablecomments{Input parameters for all three injection nozzles (in code units). 
Note that for $t=0$ these are identical, while for the time-varying nozzles the boundary conditions change over time.
Shown in columns 2-8 are the input values of jet density $\rho_{\rm j}$, ambient density $\rho_{\rm a}$, gas pressure $p$, the magnetic field z-component $B_{zc}$, 
the magnetic field pitch angle $\delta_c$, the angular velocity $\Omega_c$, and the jet Lorentz factor $\gamma_{c}$. 
The subscript $c$ denotes the respective values at the z-axis, thus i.e. at $r=0$. 
Also shown are (columns 9-14) the initial values of selected physical parameters (averaged over the nozzle), such as 
the proper sonic Mach number $\mathcal{M}_{\rm s}$, the proper Alfve\'enic Mach number $\mathcal{M}_{\rm a}$, the magnetization $\sigma$, 
the plasma $\beta$ parameter, and the extractable power $P_{\rm x}$. 
Further, we show the total densities of kinetic energy $E_{\rm kin}$, the magnetic energy $E_{\rm mag}$, and the thermal energy $E_{\rm th}$.
Columns 18-21 show the initial particle parameters such as the initial power-law index $\alpha$, minimum and maximum energy cutoffs
$\gamma_{\rm min}$ and $\gamma_{\rm max}$, respectively, and the equipartition deviation factor $\epsilon$ concerning magnetic energy and particle energy. }

\end{deluxetable*}

\subsection{Particle Injection}
\label{sec:particle_injection}
So far we have discussed how we numerically govern the dynamics of the jet propagation.
Here, we present how we implement particles into the jet, how we follow them in a Lagrangian approach,
and how we accelerate them.

Overall, we apply the particle module for PLUTO invented by \citet{vaidya2018}.
From the jet nozzle, along with the fluid material, we also inject Lagrangian macro-particles into our simulation domain, which, at each spatial point, always have the same velocity as the fluid. 

These macro-particles should be understood as ensembles of micro-particles (electrons) that evolve and flow with the jet material and subsequently 
distribute over the jet volume. 
The energy distribution of these micro particles establishes the energy spectrum of the macro-particle. 

In order to avoid ambiguity in notation, from now on in this paper we denote the macro-particle (the ensemble of micro-particles as 
\textit{particle} and the constituent micro-particles as \textit{electrons}.

For the particle injection we define a region of circular cross-section with radius $r_j$ and at a height of one pixel above the injection nozzle. 
This cross-section is divided into a cylindrical grid with $25$ uniformly distributed radial divisions (along $r$) and $36$ randomly distributed 
angular divisions (along $\phi$) with $0 < \phi < 2\pi$, {\rev that are fixed in time}. 
Here $r$ and $\phi$ are usual cylindrical coordinates. 
Hence, we get a collection of $25 \times 36$ randomly distributed points from which we inject Lagrangian macro-particles. 
This injection is done after every two (numerical) time steps {\rev so that} for a simulation run lasting 50 physical time steps, we get $\simeq 10^6$ such particles injected into the domain, 
ensuring a proper sampling of the jet material. 

Each macro-particle represents an ensemble of non-thermal electrons with an energy distribution following a power-law
\begin{equation}
\label{eq:power-law}
    \mathcal{N}(\gamma) = \mathcal{N}_0 \gamma^{-\alpha}    
\end{equation}
with a chosen {\em initial} power-law index $\alpha = 6$. 
We apply initially a power-law cut off at $\gamma_{\rm{min}} = 10^2$ and $\gamma_{\rm{max}} = 10^8$.  
Obviously, when the jet evolves, particle acceleration and non-thermal cooling will affect the cut-off energies as well
as the particle distribution evolves away from the initial power law. 
With that, $\mathcal{N}(\gamma)$ measures the number of particles per unit volume with a Lorentz factor $\gamma$, 
while $\mathcal{N}_0$ is defined by the number density of electrons $N_{\rm{e}}$  as
\begin{equation}
\label{eq:nmicro}
    \int_{\gamma_{\rm{min}}}^{\gamma_{\rm{max}}} \mathcal{N}_0 \gamma^{-\alpha} d\gamma = N_{\rm{e}}.
\end{equation}
With a chosen normalization of $m_{\rm e}c^2 = 1$, the Lorentz factor of the electron $\gamma$ then
gives us the energy of the electron, where $m_{\rm e}$ is the mass of the electron. 
Hence, in order to convert the particle energy into physical scales, the Lorentz factor of the electron $\gamma$ must be multiplied with $m_{\rm e} c^2 = 0.511$ MeV.

Finally, we quantify $N_{\rm{e}}$ by using assumption of equipartition
of energy density between magnetic field and radiating electron giving
\begin{equation}
\label{eq:equip}
    U_{\rm e} = m_{\rm e} c^2 \int_{\gamma_{\rm{min}}}^{\gamma_{\rm{max}}} \gamma \mathcal{N}(\gamma) d\gamma = \frac{B_{\rm{eq}}^2}{2} = \epsilon^2 \frac{B_{\rm{dyn}}^2}{2}
\end{equation}

where $U_{\rm e}$ is the energy density of electron and $B_{\rm{eq}}$ represents the magnetic field corresponding to the equipartition. 
The magnetic field $B_{\rm{dyn}}$ in our simulation, varies from $B_{\rm{eq}}$ by a factor $\epsilon$, which can be used to vary the ratio of the energy density 
between the magnetic field and the radiating electrons in our simulations. 
The value of $N_{\rm{e}}$ can now be calculated by substituting Equation~\ref{eq:power-law} and Equation~\ref{eq:nmicro} in Equation~\ref{eq:equip} as
\begin{equation}
    N_{\rm{e}} =  \frac{\epsilon^2}{m_e c^2} \frac{B_{\rm{dyn}}^2}{2} \left(\frac{2-\alpha}{1-\alpha}\right)
    \left( \frac{\gamma_{\rm{max}}^{1-\alpha} - \gamma_{\rm{min}}^ {1-\alpha}}{\gamma_{\rm{max}}^{2-\alpha} - \gamma_{\rm{min}}^ {2-\alpha}} \right)
\end{equation}
{\rev Since the magnetic field in the jet injection nozzle varies following the profiles mentioned in Appendix~\ref{appendix: inj_profiles}, 
we define $B_{\rm dyn}$ as the average total field strength in the injection nozzle, 
giving $B_{\rm dyn} = 6.27$}.
With a choice of $\epsilon^2 = 8 \times 10^{-5}$ in our simulations,
such that the particle energy is initially in sub-equipartition with the energy in the magnetic field,
we get $N_{\rm e} \sim 0.29$ and $\mathcal{N}_0 \sim 1.46 \times 10^{10}$. 

Subsequently, these particles are advected along with the flow and fill the outflow volume. 

\subsection{Particle Acceleration \& Evolution}
\label{sec:radiation}

{\rev After being injected at the jet base with an initial energy distribution, the particles evolve in the time and momentum space
following a Lagrangian approach in which the particles follow the fluid velocity streamlines. 
As a result, the position of the particle $\textbf{\textit{r}}_{\rm p}$ is updated with time $t$ according to the equation $d\textbf{\textit{r}}_{\rm p}/dt = \textbf{\textit{v}}_{\rm p} = \textbf{\textit{v}}_{\rm f}$, where $\textbf{\textit{v}}_{\rm p}$ is the velocity of the particle and $\textbf{\textit{v}}_{\rm f}$ is the velocity of the fluid interpolated from the underlying Eulerian grid at the position of the particle. 
It must be noted that the particles here do not have any feedback on the fluid i.e. they do not change the fluid properties (e.g. density and velocity). }

This procedure takes into account the acceleration of the particle through diffusive shock acceleration. 
Additionally, we take into account energy losses due to the adiabatic expansion of the jet, 
the synchrotron cooling of the particles by acceleration in the jet magnetic field,
and IC scattering of CMB photons by the particles. 

The DSA is often invoked to explain the presence of non-thermal particles in astrophysical systems. 
It results from repeated scattering of a particle off a coherent in-homogeneous magnetic field e.g. at a shock. 
To model DSA, we follow the novel approach developed in \citet{vaidya2018}.
Here, the Lagrangian particles are flagged when entering a shocked region, defined as the area of cells with negative velocity divergence, 
together with a  pressure gradient $\nabla p /p$ above some (normalized) threshold of 2. 
{\rev Following these conditions, the shocks in PLUTO are resolved with 3 grid cells.}

Particles are followed carefully as they travel through the shocked region, and the pre-shock and post-shock states are defined as 
the states with the minimum and maximum value of pressure, respectively. 
Based on these states, we calculate the orientation of the magnetic field with respect to the shock normal 
and the strength of the shock, which is quantified by calculating the compression ratio
\begin{equation}
\label{eq:cmpr}
    \eta = \frac{v_{\rm u}}{v_{\rm d}} = \frac{\rho_{\rm d} \gamma_{\rm f,d}}{\rho_{\rm u} \gamma_{\rm f, u}}
\end{equation}
in the shock rest frame for relativistic case (for a detailed discussion see \citealt{guess1960, lichnerowicz1976}). 
Here, the subscripts 'u' and 'd' denote the upstream and downstream values, respectively.
Using these parameters, the post-shock distribution of particles is updated as per the theory of DSA (see \citet{vaidya2018} and references therein).

{\rev Radiative losses in our simulations occur due to synchrotron process, as well as up-scattering of the surrounding CMB radiation through the IC process. 
These processes are implemented in the particle module of PLUTO by solving the time-dependent particle transport equation \citep{webb1989, vaidya2018}
\begin{equation}
\label{eqn:transport}
    \frac{d\mathcal{N}(\gamma)}{d  \tau} + \frac{\partial}{\partial \gamma} \left[ \left( -\frac{\gamma}{3} \nabla_{\mu} u_{\rm f}^{\mu} + \dot{\gamma}_l \right) \mathcal{N}\right] = -\mathcal{N} \nabla_{\mu} u_{\rm f}^{\mu}
\end{equation}
where $\tau$ is the proper time, the first term inside the round bracket represents the losses from adiabatic expansion, $\dot{\gamma}_l$ represents radiative loss due to synchrotron and IC processes, and $u_{\rm f}$ is the proper fluid velocity. 
The above equation is solved for each macro-particle advecting spatially with the fluid considering that the spectral distribution $\mathcal{N}(\gamma)$  of constituent micro-particles (i.e., electrons) is evolved at every step 
accounting for the above loss and shock acceleration processes based on local conditions interpolated from the fluid grid at particle's position.
}

%
\section{Jet Dynamics and Energetics}
\label{sec:res_dynamics}
We first discuss general features of the jet evolution before we compare specifics of
how the different jet nozzles affect the structure of the propagating jet.  

\subsection{General Jet Evolution}
\begin{figure*}[t]
    \centering
    \includegraphics[width = 0.9\linewidth]{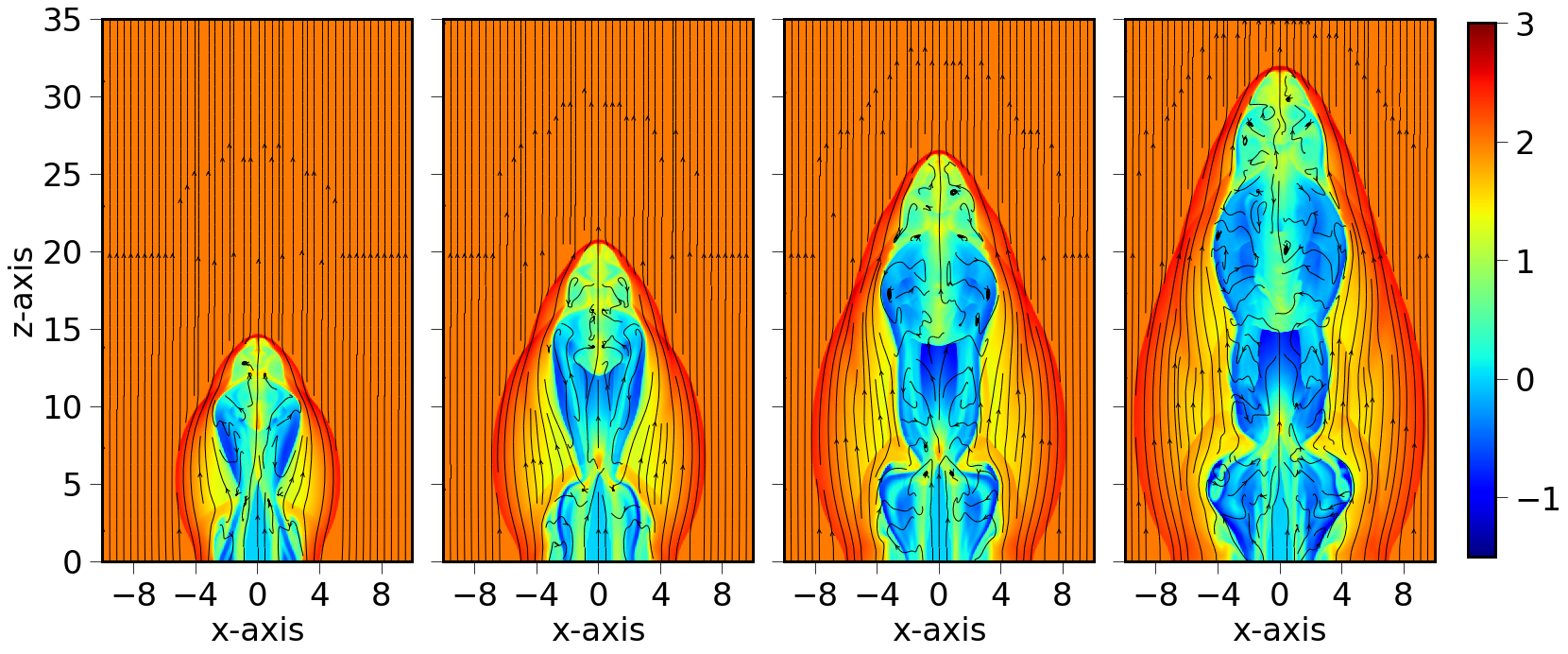}
    \includegraphics[width = 0.9\linewidth]{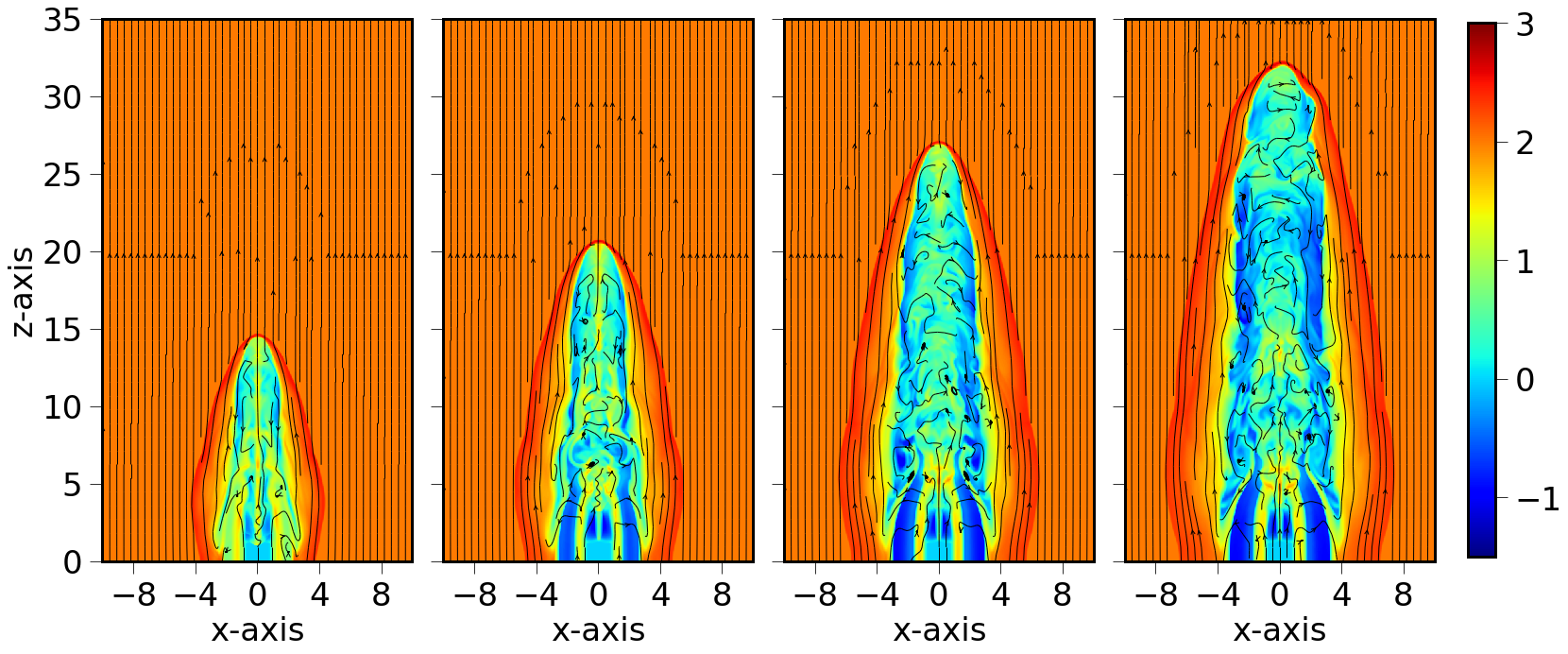}
    \includegraphics[width = 0.9\linewidth]{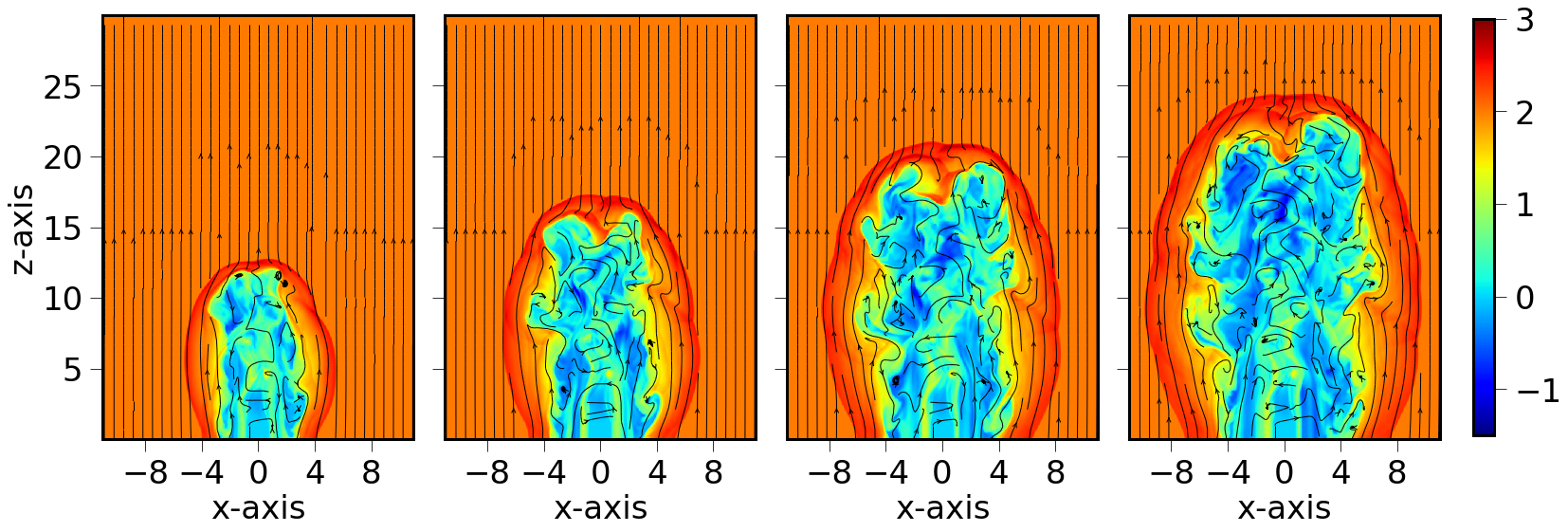}
\caption{Distribution of density (in $\log$ scale) in the domain at a time 
 (\textit{left to right}) $t =$ 20, 30, 40 and 50 (in code units) in the $x-z$ plane 
  for the simulation run \textit{std} (upper panel), \textit{var} (middle panel) and \textit{prc} (lower panel). 
  Magnetic field lines projected onto the plane are also shown (black lines).}
    \label{fig:rho_evol}
\end{figure*}

In Figure~\ref{fig:rho_evol}, we show the evolution of the dynamical structure of the jet
for the different approaches of a time-independent injection (steady nozzle, \textit{std}) in the top panel, a time-variable injection (\textit{var}) in the middle panel, and a precessing nozzle (\textit{prc}) in the bottom panel.
We display 2D slices of the 3D density distribution superimposed by (projected) magnetic field lines in the $x-z$ plane at $y=0$ at dynamical times $t= 20, \ 30, \ 40,$ and $50$ (from left to right).

\begin{figure*}[t]
    \centering
    \includegraphics[width = 0.9\linewidth]{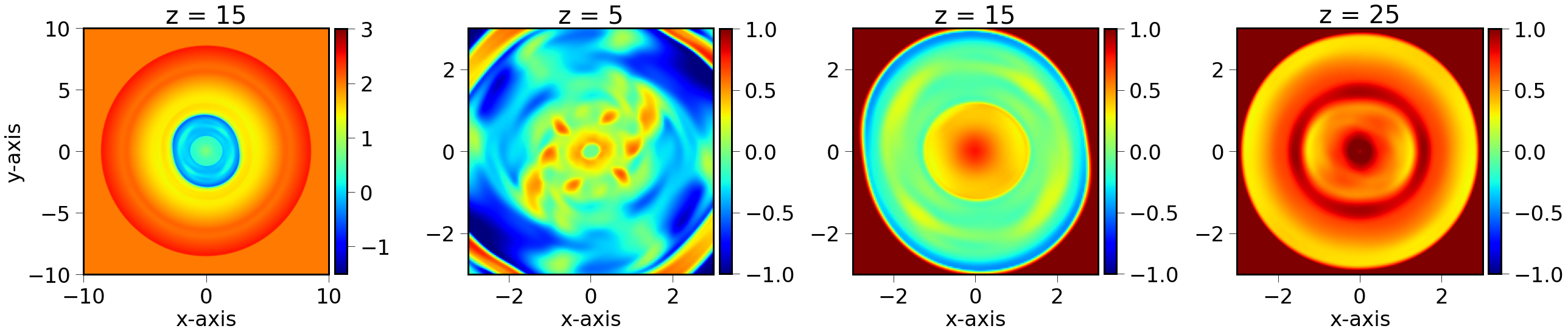}
    \caption{Cross section of the jet density distribution (in $\log$ scale) at $t =50$ in the $x-y$ plane at $z= 5, 15, 25$ (see panel title)
    for the simulation run \textit{std}. 
    Note the different sizes {\rev and different color bars} of the first and the other three panels.
    }
    \label{fig:rho_zcut}
\end{figure*}

Following the injection from the nozzle, the steady jet (\textit{std}) propagates at high speed 
in $z-$direction, filling the domain with low-density jet material before terminating at the 
jet head with a termination shock.
Subsequently, the jet inflates a low-density cocoon, thereby forming shocks as a result of 
interaction with ambient medium and also due to interaction within the jet. 
The steady jet also encounters re-collimation shocks as it evolves.

Note, however, that the strength and even the existence of these re-collimation shocks depends 
critically on the numerical resolution applied (see Appendix~\ref{appendix:resolution}),
which emphasizes the need of high-resolution simulations.
The turbulent nature of the jet is evident from the small-scale fluctuations of magnetic field 
in the jet as compared to a smoother structure around the jet.
Despite of its turbulent structure, the jet remains rather symmetric about the $z$-axis
(but see our discussion in Section~\ref{sec:lateral_structure} concerning the lateral outflow structure).

We may distinguish four dynamically distinct regions in the simulation domain.
First is the high velocity jet, producing strong shocks.
Then there is the back-flow of shocked jet gas around the jet that also carries gas entrained from the ambient medium.
Further out comes a cocoon structure with shocked ambient gas.
All this is enclosed by the original ambient gas that is not yet affected by the jet flow and remains
at rest.

An interesting feature, again a feature that is not seen in low-resolution runs, is 
the existence of a sharp discontinuity at $z \sim 15$ in the jet. 
{\rev We call this feature the \textit{steady internal shock}}. 
This shock surface starts to develop around time $t\simeq 30$ and does not evolve much with time. 
{\rev Along the jet axis, this steady internal shock is located at $z = 14.8$ at $t=50$.}
Essentially, it does not propagate {\rev much}, as long as the termination shock at the jet head is within the computational domain. 
We understand this as resulting from the interaction between the jet beam flowing forward 
and the presence of an {\em inner} back-flow\footnote{The {\em inner back-flow} denotes material 
flowing in the negative $z-$direction along the jet axis. 
This results from the deflection of jet material at the termination shock and can be seen in the  $v_z$ distribution
(for a detailed discussion see Section~\ref{sec:std_jet}).}. 
{\rev To understand quantitatively the kinematics of the steady internal shock, 
we find that between time $t = 40$ and $50$, the jet head advances from $z = 26.5$ to $z = 32.5$ i.e. a distance $\simeq 6$ (in code units). 
In the same time interval, the steady internal shock advances from $z = 13.9$ to $z = 14.8$, a distance of $0.9$ (in code units).}

Our explanation is supported by the observation that when running the simulation for longer times,
such that the termination shock at the jet head leaves the domain, there is no material 
deflected from the jet head into the original jet beam.
As a result, the {\rev steady internal} shock surface starts to evolve with time and fades away.
An interesting manifestation of this strong, steady {\rev internal} shock feature could be the formation of a knot in the emission map. 
{\rev Prime observational examples of such a steady knot feature is HST-1 in M87 \citep{biretta1999, harris2006}, and other knot-like features in AGNs discussed in \citet{hervet2016, lister2021}.}
A detailed investigation of this feature will be presented in a subsequent study.

The jet with a variable velocity (\textit{var}) forms a number of bow-shocks as it propagates, 
resulting in a skeleton-like appearance. 
These bow-shocks can be important for particle acceleration in the jet as we will investigate in 
further sections. 
This jet remains more collimated than the jet with steady injection, but shows a similar linear
extent along the propagation direction. 

The precessing jet (\textit{prc}) is, obviously, highly asymmetric.
Essentially, it shows a smaller linear extent in the propagation direction as compared to the 
steady jet at the same time, a natural consequence of the smaller momentum in this direction. 
We note, however, that the magnitude of the total velocity in both cases (precessing and steady jet)
is same and hence, we inject same total kinetic energy density $\rho v^2$ in the two cases.  
As a result of precessing velocity vector of the injected material, we see multiple {\em finger-like} extensions of the jet head.
Note that this is a truly 3D feature, while the figures show slices of the 3D structure.
This can also result in enhanced interaction with the ambient medium, which can result in more shocks and turbulence as we discuss in the next section.

We already mentioned the high degree of symmetry we observe in the jets injected without precession. 
This symmetry shows up despite the 3D treatment of the simulation and also despite the
application of Cartesian coordinates, altogether demonstrating the quality of our numerical setup. 
{\rev Note however, that there are locations around the jet where the axial symmetry is partly broke (see discussion in next section).}

\subsection{Lateral Jet Structure}
\label{sec:lateral_structure}
We further investigate the lateral structure of the jet by showing slices of the density
distribution across the jet thus in the $x-y$ plane 
along different distances along the jet and also for a chosen time $t=50$ (see Fig.~\ref{fig:rho_zcut}).

When we look at the full extent of the outflow structure at $z=15$ (left panel), wee see a low-density 
(in blue) structure in the center -- the high velocity jet -- surrounded by a 
structure of intermediate density (yellow), that is moving with lower (and negative) speed,
which we interpret as {\em back-flow} of shocked jet material.
Both are surrounded by a high density
and low velocity (in arbitrary direction) cocoon {\rev (in red)} that contains
shocked material from the ambient medium.
Outside the cocoon we find the uniform ambient medium (in orange) -- still in its initial state.

The cocoon shows a perfectly circular shape, implying that we have sufficient resolution to avoid instabilities generated as a 
result of injecting a cylindrical jet in a Cartesian grid.
The other panels show a zoomed-in version of the density distribution at different values of $z$,
all at the same time.
These panels resolve the innermost jet structure close to the jet axis.

At $z=5$, thus close to the injection nozzle, we see features resulting from active mixing of jet material with the back 
flow material. 
At the very center we see the original jet beam with density $\rho = 1$ (i.e. $\log \rho = 0$). 
This is surrounded by a denser region formed as a result of shocks at the contact discontinuity at the jet beam boundary 
and the back-flowing jet material.  

Note the elliptical region of small substructures that has a factor ten time higher density as compared to the jet beam .
Interestingly the orientation of this ellipse is not aligned with the numerical grid.
Also the number of these features is six, and not four as maybe expected from the 
Cartesian grid.
We may understand this substructure as braids.
Just outside these elliptically aligned braids we see two features ({"}red{"}) that belong to a
spiral structure.
Indeed, further out the lower density material (greenish) seems also aligned in a spiral shape.
Even further out, at $r\simeq 3$, again six features appear, in similar order as the inner
braids.
The physical reason of this ordered substructure is yet unknown to us.
In principle, however, we may think about this as longitudinal re-collimation shock forming
a braided structure. 
{\rev The exact origin of this ordered substructure may also be related on the profiles of variables injected from the nozzle,
in particular from the profile of the jet rotation.
However, we note that these braids are located {\em outside} of the jet stream (even at the location of the re-collimation shock),
where there can hardly be shear between the jet rotation and the surrounding gas.
Interestingly, while these braids break the toroidal symmetry of the outflow, their alignment in $\phi$-direction shows still a degree of symmetry.
Furthermore, further downstream, the jet flow recovers again a good degree of axisymmetry. }

Further downstream, {\rev we show the zoomed-in version of the first plot of Figure~\ref{fig:rho_zcut} at $z=15$. We see} a denser region in the center, implying that we 
have captured the shocked area just above the sharp gradient in density and pressure {\rev at the steady internal shock}. 
The substructure of two times six braids has disappeared, the jet forms a ring-like structure
of rings of different density, with the central jet being densest.

Even further downstream from the nozzle, at $z=25$, the jet approaches the jet head
where we observe a high density due to the presence of the termination shock.

\begin{figure*}[t]
    \centering
    \includegraphics[width = \linewidth]{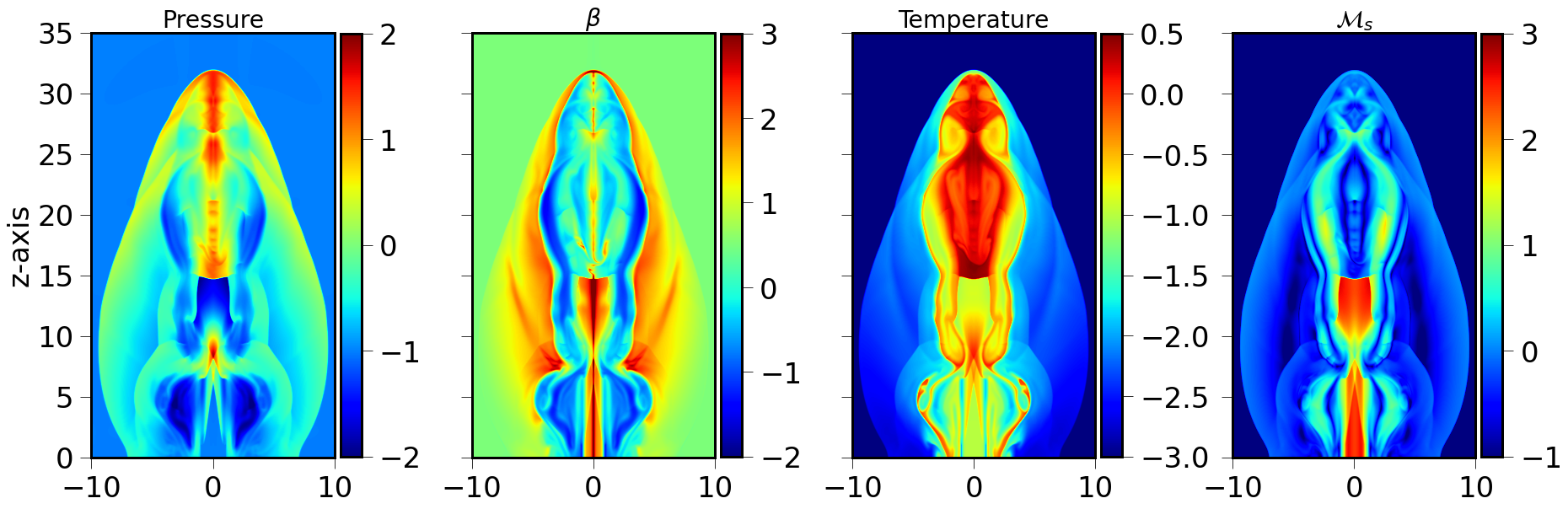}
    \includegraphics[width = \linewidth]{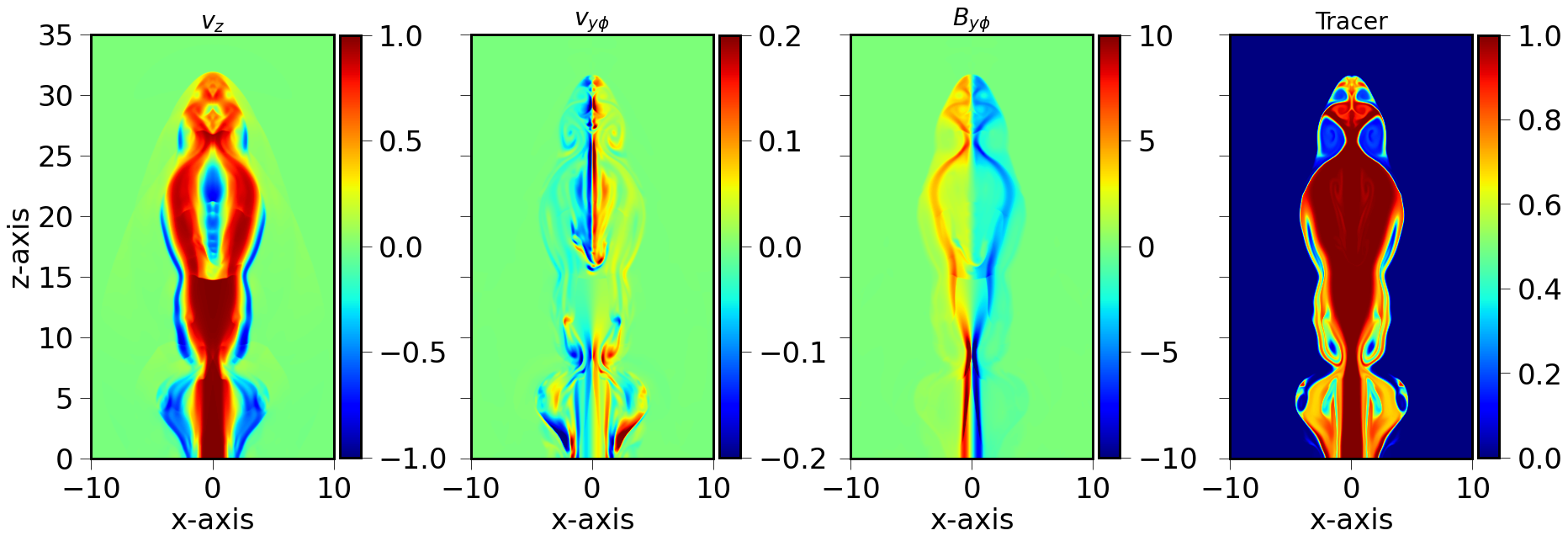}
    \caption{\textit{Top: }Distribution of (\textit{left to right}) 
     pressure (in $\log$ scale), 
     plasma $\beta$ (in $\log$ scale), 
     temperature $\Theta$ (in $\log$ scale), and 
     proper sonic Mach number $\mathcal{M}_s$ (in $\log$ scale) in the domain at a time  $t =50$ (in code units) 
     in the $x-z$ plane for the simulation run \textit{std}.
    \textit{Bottom: }Distribution of (\textit{left to right}) 
     $v_z$, the 
     $y-$component of $v_{\phi}$, the 
     $y-$component of $B_{\phi}$, and the
     passive tracer {$tr$} 
     in the domain at a time $t =50$ (in code units)
     in the $x-z$ plane for the simulation run \textit{std}.
     }
    \label{fig:std_plots}
\end{figure*}

\subsection{Further Jet Diagnostics}
\label{sec:std_jet}
In order to see how the other dynamical properties of the jet vary over the domain, we show 2D slices of various jet variables at time $t=50$,
again concentrating on the steady jet nozzle, simulation \textit{std} (see Fig~\ref{fig:std_plots}).
The pressure distribution shows, like the density distribution, a number of re-collimation shocks in the jet, as well as 
the {\rev strong steady internal shock} feature at 
$z \sim 15$, and the termination shock near the jet head. 

These features are also visible in the plasma $\beta = 2p/B^2$, the temperature $T$, and the proper sonic Mach number $\mathcal{M}_{\rm s}$. 
We show the toroidal components of velocity and magnetic field, by plotting the $y-$component of $v_{\phi}$ and $B_{\phi}$. 
Through this, we can visualize $v_{\phi}$ and $B_{\phi}$ coming out (in blue), and going in (red) the $x-z$ plane. 
We note that by this we also see the axisymmetric nature of the fields demonstrated, strongly indicating that we have applied a proper 
resolution when injecting a cylindrical jet into a Cartesian domain. 
We also note the opposite signs of toroidal velocity and magnetic fields in the injection profiles (Equation~\ref{eq:b_phi}).
This reversal of sign is a typical consequence of jet launching and acceleration models \citep{BP1982, ferreira1995, zanni2007}. 

The diagram for the vertical velocity $v_z$ shows an interesting feature of jet dynamics. 
As the jet develops, it expands and eventually undergoes a {\em bifurcation} at $z \sim 15$.
This results from the interaction with back-flowing material that was reflected by the jet head. 
Subsequently, the propagating jet material flows around this inner back-flowing material.
Note that the existence of such a feature is not straightforward, as the back-flowing material can be expected to
bypass the jet material {\em around} the jet.
Here, the jet reflection is so strong, and of such a reflection angle that it comes back right towards the jet flow.
We believe that the reason for this flow structure is the re-collimation shock right behind the jet head,
leading to a reflection along the jet axis. 

After all, this particular flow feature leads to the formation of the {\rev strong steady internal shock} structure that we see in the distribution 
of various physical variables.
The essential condition for this {\rev steady internal shock} to form, we think is the existence of the strong shock at the jet head.
In fact, if we let the jet head move out of our computational domain, such that there is no jet head to reflect the jet material 
(and hence, no back-flow), the strong {\rev steady internal shock} fades away.

The strong head-on collision between the jet material and the inner back flow does not
only provide a site for strong particle acceleration (see below), but also for neutrino production
(see e.g. \citealt{britzen2019a} for an observationally motivated scenario). 
From our modeling we indeed expect hadronic material to be entrained into the strong inner back-flow, 
as it carries shocked material.

Additionally, we show the distribution of a tracer for the origin of the material, where a value of unity tracks the material that originates 
from the jet nozzle, and a value of zero for material that does not belong to the jet, rather to the initial ambient medium.
This enables us in principle to disentangle the jet material from the ambient gas. 
The intermediate values of the tracer, $0 < tr < 1$, basically quantify the mixing state of two media. 
We see that as the jet propagates, it is expanding, also as a result of encountering the inner back-flowing material at $z \sim 15$. 
Note that the back-flowing material (with negative velocity $v_z$) is indeed a mixture of jet material and ambient gas and
and hence has a tracer value $<1$.

\subsection{Jet Nozzles}
Before we compare the particle acceleration and non-thermal cooling processes 
in our simulations applying different jet nozzles,
here we first investigate the differences in the dynamical features of jet propagation that are governed by
the nozzle physics.
Eventually, the jet dynamics will directly determine the acceleration and radiation properties of the jet. 

\begin{figure*}[t]
    \centering
    \includegraphics[width = \linewidth]{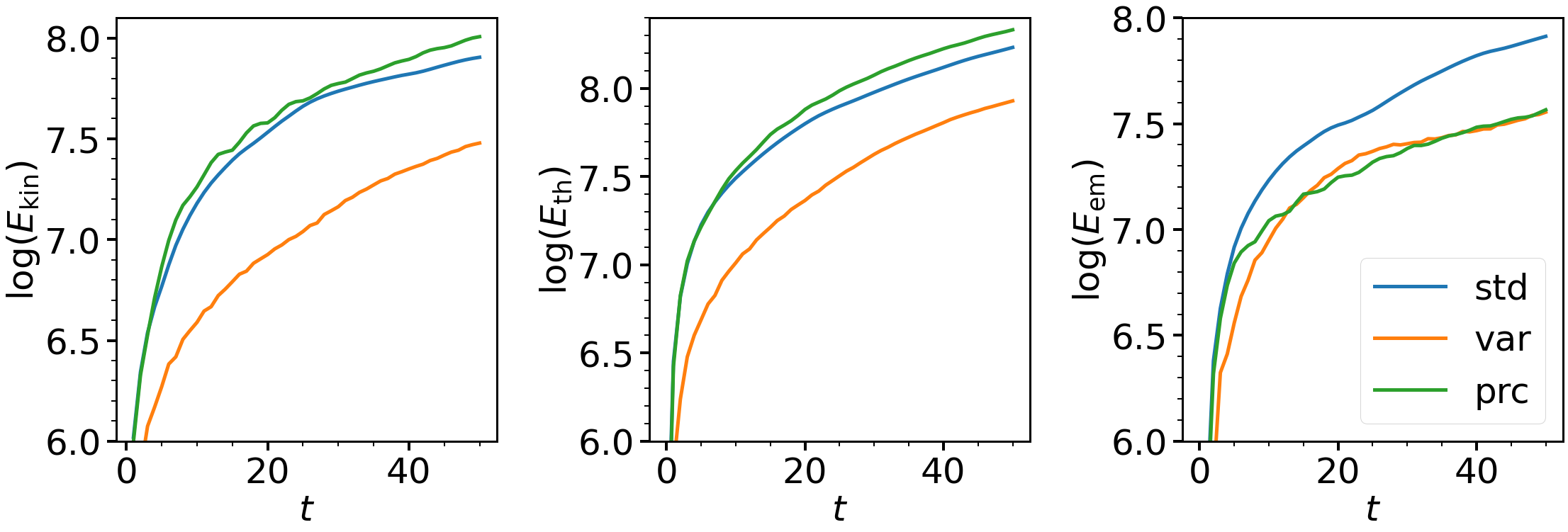}
    \caption{Time-evolution of global outflow energetics.    
     Shown are integrated values of
     the kinetic energy density $E_{\rm kin}$, 
     the thermal energy density $E_{\rm th}$, and 
     the electromagnetic energy density $E_{\rm em}$  ({\em from left to right}). 
     We consider computational values beyond $t=1$ and integrated from above $z=2$.
     }
    \label{fig:int_evol}
\end{figure*}

To this end, we consider the contribution of various dynamical parameters only from the jet-cocoon structure and
ignore the ambient medium. 
For this purpose, we define the jet-cocoon region as being composed by the grid cells where the speed $v > 0$. 
Then, the value of a certain parameter at a particular time is calculated by summing up the contribution from all the grid cells in 
the jet-cocoon region at that time. 
In Figure~\ref{fig:int_evol} we show for all the three nozzles, the time evolution of {\em total} energy density in different
energy channels, {\em viz} 
the kinetic energy density $E_{\rm kin}$, 
the thermal energy density $E_{\rm th}$ and 
the electromagnetic energy density $E_{em}$, each defined as in equation~\ref{eq:E_x}.

We first see that the outflow dynamics produced by the {\em std} and {\em prc} setup nozzles show similar kinetic energy density, 
whereas the \textit{var} setup nozzle evolves the outflow into a state of somewhat lower kinetic energy density. 
This is expected as the kinetic energy density we inject depends on the velocity (or Lorentz factor) injection of the jet, that is 
on average lower in the variable injection nozzle {\em var} as compared to the other two nozzles, which have the same velocity injection. 
Comparing the injection nozzles {\em std} and {\em prc}, both of which have the same kinetic energy density at initial times, we find 
that the precessing outflow has a higher kinetic energy at later times.
Here, the jet is constantly changing its direction, {\rev and the interaction with the ambient medium is weaker.
As a result, less of kinetic energy is converted to other energy channels for the case of the precessing outflow.} 

For the magnetic energy density we find that the steady jet approach leads to the highest level,
followed by the variable and the precessing jet.
This is a consequence of our injection profiles, as the magnetic field which is injected along the jet depends on the jet velocity 
(see Equation~\ref{eq:b_phi}).
Thus, the variable jet outflow is expected to be less magnetically energized than the steady jet.

The precessing jet on the other hand, produces and contains jetlets\footnote{With {\em jetlets} we denote the jet portions that are injected from the nozzle at each time.
These jetlets can also be understood as small volumes of material traveling together with the same bulk velocity.} moving in different directions.
These jetlets, on interacting, may lead to (numerical) reconnection of magnetic field. 
This is interesting from the point of energetics, as this will also lead to (numerical) heating of the jet.
Due to the high resolution we apply, numerical diffusion and reconnection does play a minor role only.
However, this energy will have in principle implications on the particle acceleration in such jets. 
In our approach of ideal MHD we cannot not model physical magnetic reconnection. 

Another interesting point to note here is that although the steady and the variable jet are in equipartition with respect to the kinetic, thermal and electromagnetic 
energy densities, 
the precessing jet with similar injection as the steady jet is less magnetically dominant than the other jets we discuss. 

\subsection{Outflow Turbulence Level}
\label{sec:turbulence}
In the approach of the present paper, acceleration of particles is achieved by shocks.
Shocks are typically generated by a  misaligned velocity field.
Such velocity fields are a typical signature of turbulence.
From Figure~\ref{fig:rho_evol}, we (literally) see that the level of turbulence looks different for the different jet 
nozzles we apply.
We thus expect that the various nozzles we consider in this work drive turbulence in the domain differently based on their dynamics. 
We now want to quantify the level of turbulence generated by the various injection nozzles.
 
In this context, we define the level of turbulence in  a variable $X$ at a certain grid cell 'i' located at (x, y, z) as
\begin{equation}
    {\delta} X^{\rm T}_i = | X_i(x,y,z) - \overline{X}_i | 
\end{equation}
where $\overline{X}_i$ is the mean value of the variable $X$ over all grid cells inside a cube of $20^3$ grid cells, centered at $(x,y,z)$.  
Note that for calculating the local turbulence level ${\delta} X^{\rm T}_i$, 
we only consider grid cells $i$ where the tracer $tr > 0.1$.
So, we do not take into account the grid cells that are in the ambient medium for calculating turbulence as well as for averaging. 
Also, we consider grid cells only above $z=2$, thus above the injection nozzle only.
As a measure of the total level of turbulence of the jet we then sum over all local measures of turbulence
$X^{\rm T} = \Sigma_i {\delta} X^{\rm T}_i$ (independently for all components $x,y,z$ if $X$ represents a vector field).

We can now compare the turbulence level in the velocity and the magnetic field for the different jet nozzles 
by calculating spatial averages of $v^{\rm T}/v$ and $B^{\rm T}/B$, respectively. 
We find that both, the variable jet nozzle and the precessing jet, drive a higher level of turbulence in the velocity 
with an average $v^{\rm T}/v \simeq 0.21$,
while for the steady jet we find average $v^{\rm T}/v \simeq 0.15$. 
Similarly, for the turbulence level in magnetic field we find $B^{\rm T}/B \simeq 0.10, 0.17, 0.33$ for the steady, variable, and precessing jet, respectively.

Another way of quantifying the level of turbulence is to compare the different energy channels {\em viz} the kinetic, thermal and electromagnetic energy densities 
$E_{\rm kin}^{\rm T}$, $E_{\rm th}^{\rm T}$, and $E_{\rm em}^{\rm T}$, respectively, {\rev as defined in equation~\ref{eq:E_x}}.
We show the exact values of these terms in Table~\ref{tbl:output}.
These can then be compared to the corresponding (integrated) energy terms (see Equation~\ref{eq:E_x}), 
which we also show in Table ~\ref{tbl:output}, providing the fraction of the total energy transformed 
into turbulent energy. 
After all, we expect that different nozzles govern a different fraction of the injected energy that is turned into turbulent energy.

Computing these fractions we find that $\simeq 1.12\%$ of the injected kinetic energy is turned into turbulent kinetic energy for the {\em std} nozzle setup. 
Similarly, the {\em var} and {\em prc} nozzles turn $\simeq 1.81\%$ and $1.12\%$ of the injected kinetic energy into the turbulent kinetic energy, respectively. 
On the other hand, the {\em std}, {\em var} and {\em prc} nozzle convert $\simeq 1.62\%$, $0.89\%$ and $1.71\%$ of their injected magnetic energy into the turbulent magnetic energy, 
and $\simeq 0.23\%$, $0.22\%$ and $1.26\%$ of the injected thermal energy into the turbulent thermal energy, respectively.
From this comparison, we conclude that the time-variation of the jet injection, in particular the precessing jet, leads to enhanced level of turbulence in the domain.

As a turbulent jet will lead to the formation of more number of shocks, turbulence has an implication on the 
acceleration of particles as well. 
Thus, from investigating the outflow turbulence levels we predict that the precessing jet will be most efficient in accelerating particles. 

We show values of some characteristic dynamical output parameters at time $t=50$ for the different injection setups, along with the parameters connected 
to particle acceleration (see next section) in Table~\ref{tbl:output}.

\begin{deluxetable*}{ccccccccccccccccccc}
\tabletypesize{\scriptsize}
\tablewidth{0pt}
\tablecaption{Output Characteristic Values \label{tbl:output}}
\tablehead{
\colhead{Setup} & \colhead{$\overline{\mathcal{M}}_{\rm s}$} & \colhead{$\overline{\mathcal{M}}_{\rm a}$} & \colhead{$\overline{\sigma}$} & \colhead{$\overline{\beta}$} & \colhead{$\overline{\gamma}$} & \colhead{$P_{\rm x}$} & \colhead{$V$} & \colhead{$E_{\rm tot}^V$} & \colhead{$E_{\rm kin}$} & \colhead{$E_{\rm th}$} & \colhead{$E_{\rm em}$} & \colhead{$E_{\rm kin}^{\rm T}$} & \colhead{$E_{\rm th}^{\rm T}$} & \colhead{$E_{\rm em}^{\rm T}$} & $\overline{B^{T}/B}$ &\colhead{$E_{\rm p}^{\rm ef}$} & \colhead{$\gamma_{\rm min}$} & \colhead{$\gamma_{\rm max}$} \\
\colhead{} & \colhead{} & \colhead{} & \colhead{} & \colhead{} & \colhead{} & \colhead{} & \colhead{} & \colhead{$\times 10^8$} & \colhead{$\times 10^7$} & \colhead{$\times 10^7$}  & \colhead{$\times 10^7$} & \colhead{$\times 10^5$} & \colhead{$\times 10^5$} & \colhead{$\times 10^5$} & \colhead{} & \colhead{$\times 10^3$} & \colhead{($\log$)} & \colhead{($\log$)}
}
\colnumbers
\startdata
{\em std} & 2.37 & 2.79 & 1.65 & 13.4 & 1.69 & 418 & 5785 & 3.33 & 5.08 & 9.93 & 7.77 & 5.72 & 2.25 & 12.6 & 0.10 & 1.44 & 0.79 & 8.52 \\
{\em var} & 1.38 & 1.32 & 1.54 & 2.89 & 1.29 & 194 & 3876 & 1.55 & 1.90 & 5.27 & 3.43 & 3.45 & 1.15 & 3.05 & 0.17 & 0.95 & 0.35 & 8.51 \\
{\em prc} & 1.49 & 4.26 & 0.38 & 49.3 & 1.31 & 443 & 5903 & 3.54 & 5.98 & 11.6 & 3.33 & 6.71 & 14.6 & 5.70 & 0.33 & 8.93 & -0.12 & 9.07 \\
\enddata    
\tablecomments{Output parameters of simulation runs for the steady outflow setup \textit{std}, the variable outflow setup \textit{var} and the precessing outflow setup \textit{prc}(in code units) at $t=50$. 
Shown are the average values for 
the proper sonic Mach number $\overline{\mathcal{M}}_{\rm s}$, 
the proper Alfv\'enic Mach number $\overline{\mathcal{M}}_{\rm a}$, 
the magnetization $\sigma$, 
the plasma $\beta$, and
the jet Lorentz factor $\gamma$.
This is followed by 
the extractable power $P_{\rm x} \equiv E_{\rm x} V / t$, 
the outflow volume $V$ (for cells where $v>0$), 
the total energy density $E_{\rm tot}^{\rm V}$ over the whole outflow volume,
the integrated values of 
the kinetic energy density $E_{\rm kin}$, 
the thermal energy density $E_{\rm th}$, 
the electromagnetic energy density $E_{\rm em}$, and 
the turbulent kinetic, thermal and electromagnetic energy densities $E_{\rm kin}$, $E_{\rm th}$, and $E_{\rm em}$, respectively. 
The averaging and the integration is done for the cells where the tracer \textit{tr} $>0.1$. 
We also show for all the particles in the domain their total effective energy $E_{\rm p}^{\rm ef}$ (i.e. total particle energy above $\gamma = 10^3$), 
their minimum Lorentz factor $\gamma_{\rm min}$, and 
their maximum Lorentz factor $\gamma_{\rm max}$. 
}

\end{deluxetable*}

\section{Particle Acceleration}
\label{sec:res_particles}
In our simulations we inject Lagrangian macro-particles at the base of the jet which follow the flow of the fluid.

\subsection{Propagation of Particles}
\begin{figure*}[t]
    \centering
    \includegraphics[width = 0.25\linewidth]{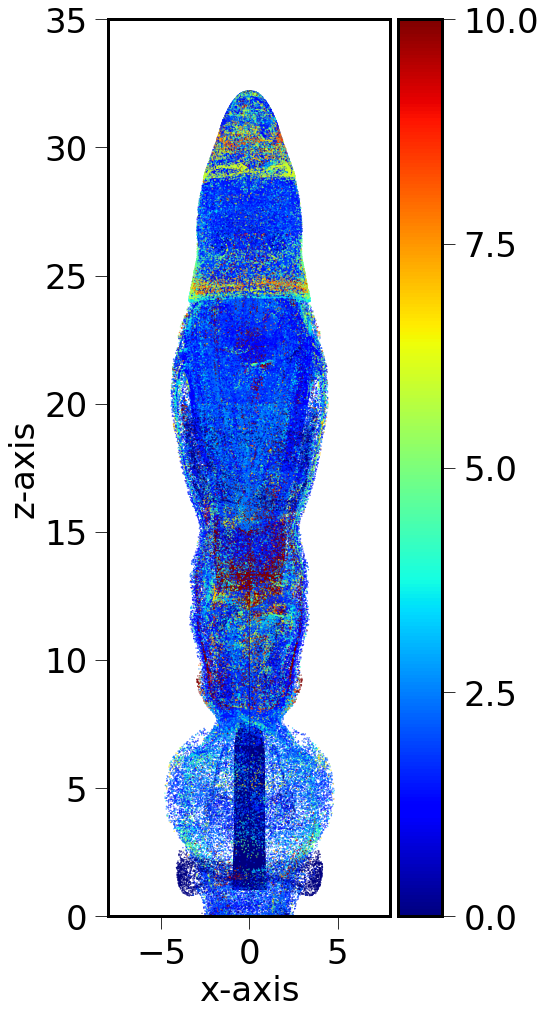}
    \includegraphics[width = 0.25\linewidth]{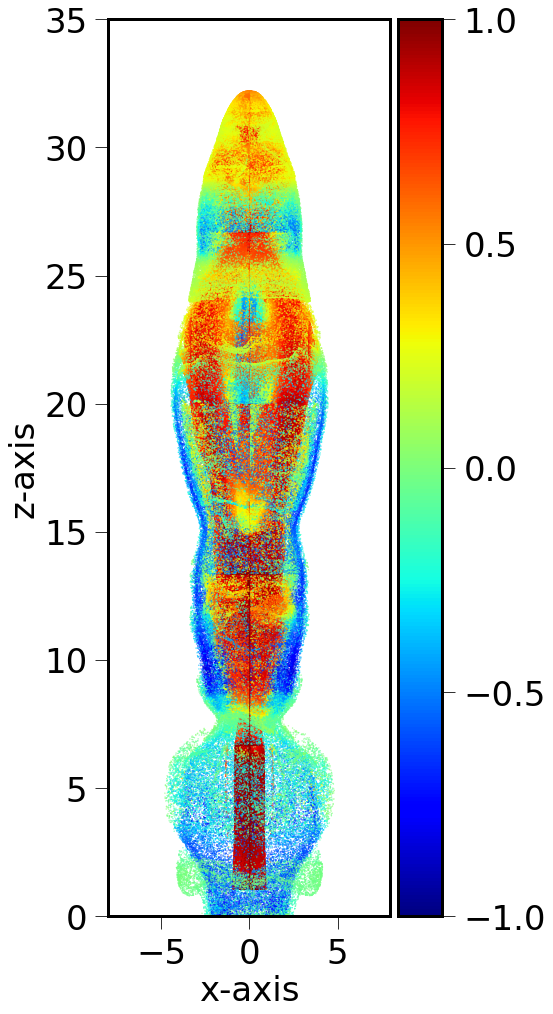}
    \includegraphics[width = 0.31\linewidth]{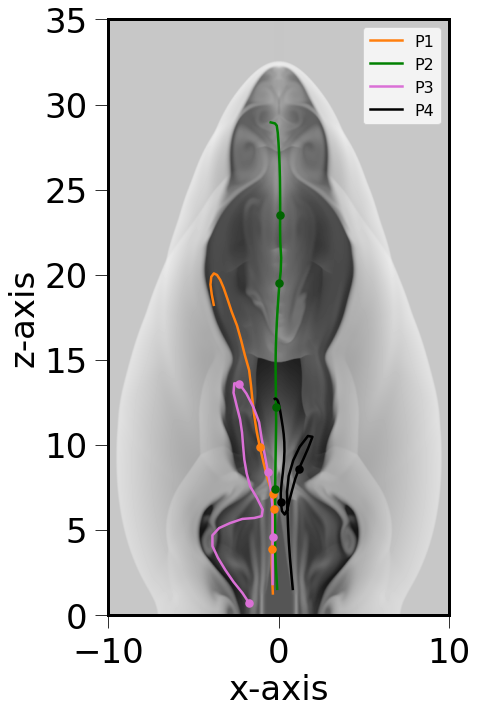}
    \caption{Distribution of particles with different compression ratio (left) and shock speed ($v_z-$component) (middle) at a time $t =50$. 
    Right panel shows example trajectories of five representative particles from injection till $t=50$,  
    with the background showing the density distribution for better visualization. 
    The colored dots on the trajectories represent the locations where the particle encounters shock.}
    \label{fig:part_dist}
\end{figure*}

In Fig.~\ref{fig:part_dist} we show the distribution of these particles in the jet for the simulation run \textit{std} at time $t=50$. 
All the particles injected are projected on the $x-z$ plane, but in reality follow a 3D distribution.
In addition, we color code these particles denoting the compression ratio $\eta$ they have experienced (in the left panel)
and the shock velocity in $z-$direction at that location (in the middle panel). 

As can be seen, particles just below the strong steady {\rev internal} shock at $z=15$ have a high {\rev compression ratio} $\eta$, suggesting the presence of very strong shocks
at this position. 
The distribution of the vertical velocity $v_z$ for the particles follows exactly that for the jet material.
Obviously this is expected for Lagrangian particles.
Hence, we see a forward going jet with positive $v_z$ (in red) along with the back-flow where particles have a negative $v_z$ (in blue). 

Figure~\ref{fig:part_dist} shows also the trajectories of few selected particles in the jet.
We may follow their path from injection evolving over time.
It can be seen that after injection at the jet base, the particles interact with the local fluid at their present 
location.
Depending on the interaction, a particle may traverse over the whole jet length or may end up in the jet back flow or cocoon 
upon reflecting off a shock.

When a particle enters a shock, the strength of the shock is given by its compression ratio $\eta$ as calculated by 
Equation~\ref{eq:cmpr}. 
When a particle leaves the shock, this value of $\eta$ is retained by the particle until it meets another shock with 
a different compression ratio.

\begin{figure*}[t]
    \centering
    \includegraphics[width = 0.45\linewidth]{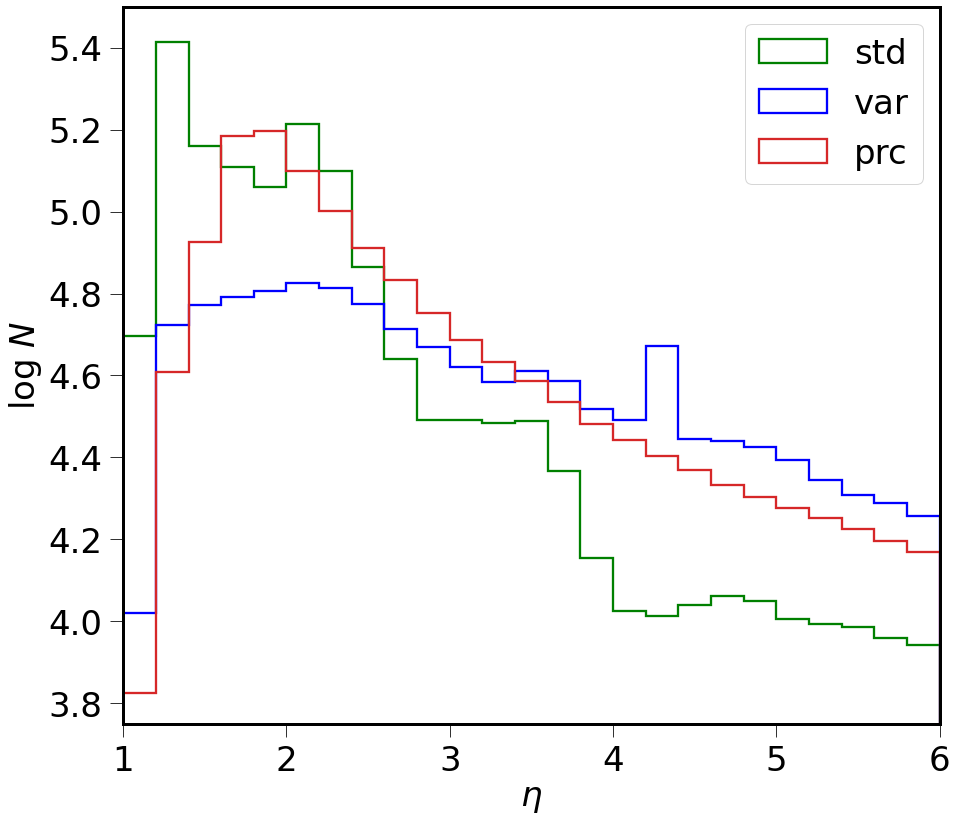}
    \includegraphics[width = 0.435\linewidth]{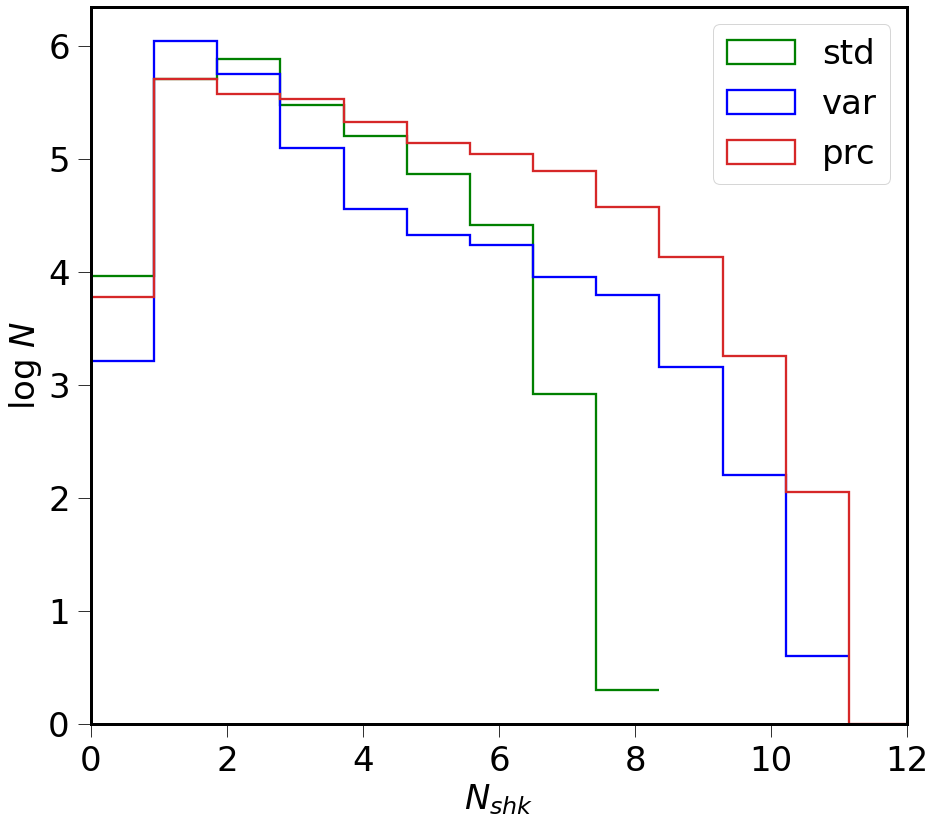}
    \caption{Histograms of number of particles \textit{N} with different compression ratio $\eta$ (left) 
     and total number of shocks $N_{\rm shk}$ (right) for the simulation run 
     \textit{std} (green), \textit{var} (blue) and \textit{prc} (red) at a time $t =50$ (in code units).}
    \label{fig:hist_cmpr}
\end{figure*}

Figure~\ref{fig:hist_cmpr} shows the statistics of the compression ratio $\eta$ at time $t=50$ over the whole 
numerical domain for simulations applying different injection nozzles. 
Note that we detect compression ratios beyond the theoretical limit of 4 for RMHD shocks. 
{\rev We note here that this classical limit of 4 for MHD shocks is enhanced by the Lorentz factor in case of RMHD shocks and can have large values \citep{blandord1976}.
However, the compression ratio $\eta$ calculated in the shock rest frame, as is done in our study for a particular shock rest frame (Normal Incidence Frame), is limited by a values of 4 \citep{guess1960, appl1988, kirk1999, summerlin2012}.  

The number of shocks with compression ratio $\eta > 4$ in our simulations} is low (<10\%) and primarily arises due to numerical artifact when the macro-particle passes through a complex network of turbulent/interacting shocks in a short 
time and is unable to resolve multiple shocks. 
As the fraction of such particles is much less than those that sample the shocks accurately, such an update will have no effect on the quantitative comparison of compression 
ratio of shocked particles from different nozzles as described below. 

We see that the variable jet nozzle {\em var} produces more strong shocks, compared to the precessing nozzle {\em prc}, which 
is similar to {\em var} concerning strong shocks.
This can be explained as being the result of different dynamics of two injection nozzle setups. 
In the {\em var} nozzle setup, the slow moving jet material injected at an earlier time is repeatedly hit by the fast moving material injected at later time during various jet injection cycles. 
As a result, there are more number of such {\em back-on} collisions in the {\em var} setup as compared to the {\em prc} setup, where the injected jet continuously changes the direction
of its motion. 
This results in more of the stronger shocks being formed in the {\em var} setup.
On the other hand, the standard injection {\em std} simulation has shocks comparatively weaker than for the other two nozzles as a result of its time-independent injection.
However, note that the number of shocks $N_{\rm shk}$ a particle has encountered along its path is not represented by the compression ratio $\eta$,
but by the number of times the $\eta$ has changed its value.

In Figure~\ref{fig:hist_cmpr} we show the histogram of the number of shocks the particles have encountered.
We see that although the precessing nozzle {\em prc} leads to somewhat weaker shocks than the variable nozzle {\em var}, 
the overall number of shocks is substantially larger.
In fact, the total number of shocked particles in the {\em var} setup is even lower than for the {\em std} setup.

Since both, the strength of the shock and also the number of shocked particles, are crucial in regard to the energetics of the particles, 
it is interesting to study which of these two is dominant.
This can be investigated by comparing the particle spectra in different jets, as we will discuss in the next section.

\subsection{Particle Energetics}
The Lagrangian macro-particles (ensemble of electrons) which are injected at the jet base follow a power-law spectrum for the constituent 
electrons given by Equation \ref{eq:power-law}. 
As the particles evolve with time (along their path), they may encounter shocks and become accelerated depending on the compression ratio they experience and the orientation of magnetic 
field with respect to the shock normal \citep[see][]{vaidya2018}. 
It should be noted that spectra of particles  with compression ratio beyond a value of 4 ($< 10$\% of all particles) is updated assuming an asymptotic limit of the power-law index $\alpha = 2.23$
for ultra-relativistic shocks \citep[see e.g.][]{kirk2000, huang2023}.

In addition to being accelerated while crossing shocks, the particles cool down due to various processes discussed in section~\ref{sec:radiation}. 
As a result, the overall spectrum of injected particles changes over time, from the initial steep power law to an energy spectrum indicating particle energy beyond and below
the initial cut-off energies and with varying slope. 

\begin{figure}
    \centering
    \includegraphics[width = \linewidth]{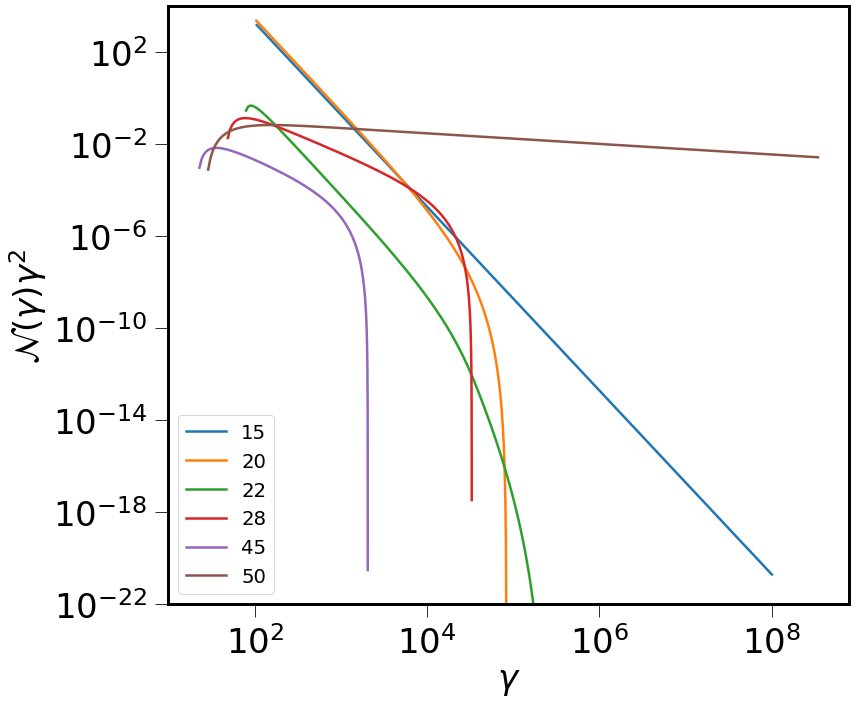}
    \caption{Energy spectrum of a particle (labeled P3 in the right panel of Fig.~\ref{fig:part_dist}) at different times (as mentioned in code units) 
    for simulation run \textit{std}.}
    \label{fig:part_spec}
\end{figure}

In Figure~\ref{fig:part_spec}
we show the spectrum of one example particle (labeled P3 in the right panel of Fig.~\ref{fig:part_dist}) for some simulation 
time steps.
A good presentation of the energy spectra is to plot the particle energy density distribution $\mathcal{N}(\gamma) \gamma^2$ as a function 
of particle energy $\gamma$.

When the particle is injected at time $t \simeq 15$, it has the initial spectral slope $\alpha = 6$ and a maximum (thermal) Lorentz
factor $\gamma_{\rm max} = 10^8$.
Note that the jet bulk motion (in z-direction) injected has a (maximum) Lorentz factor $\gamma_{\rm f}$ of 10. 

As the particle evolves, following the fluid, it cools down - with higher energy electrons of the macro particle cooling faster than the lower 
energy electrons.
As a result, we see the high-energy tail of initial spectrum decaying and a subsequent decrease of $\gamma_{\rm max}$ during $t=20$. 

Before $t \simeq 22$, the particle encounters\footnote{We do not know the exact times when the particle crosses the shock between two data dumps.
This is computed following the numerical time steps, while our data dumps follow pre-defined physical time intervals. 
We do, however, know the last time when the particle crosses the shock before a data dump.} 
a shock leading to a change in the slope of the spectrum and an increase in $\gamma_{\rm max}$ showing 
acceleration.
The particle again enters a shock region at $t \simeq 28$ and $45$, again modifying the spectral slope. 
However, as a consequence of rapid cooling of high-energy electrons, overall we see a decrease in the maximum Lorentz factor $\gamma_{\rm max}$.
Finally, at $t \simeq 50$, the particle encounters a strong shock resulting in a flat spectrum and a sudden increase in $\gamma_{\rm max}$. 

\begin{figure}[t]
    \centering
    \includegraphics[width = 1.0\linewidth]{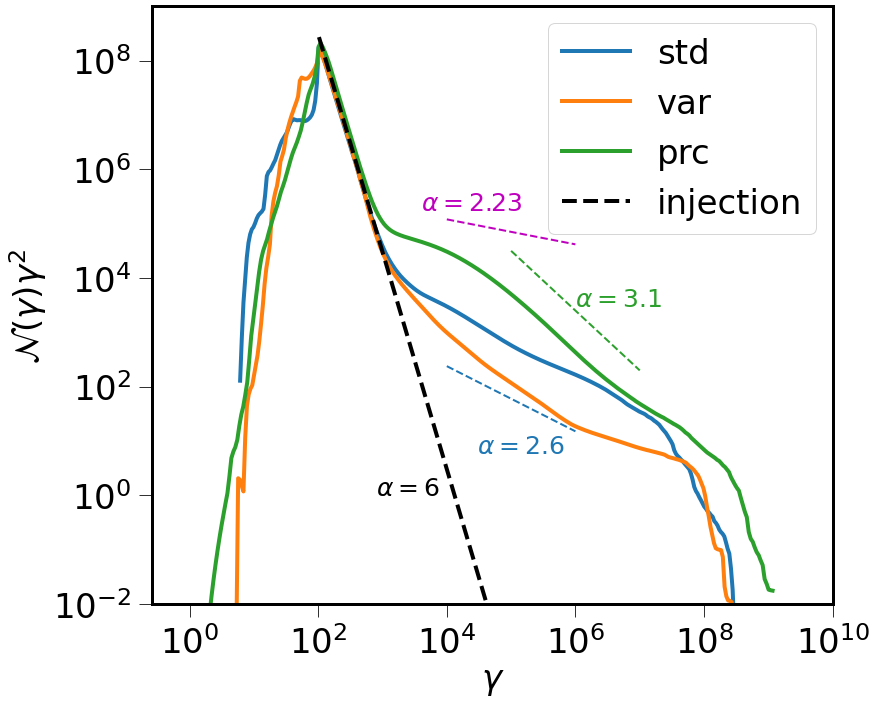}
    \caption{Energy spectrum of all particles in the domain for the simulation run \textit{std}, \textit{var} and \textit{prc} at $t =50$. 
    The injected spectrum for each particle normalized to the total number of particles at $t=50$ is shown by the black dashed line. 
    The blue and green dashed lines show the slope of spectra with power-law index $\alpha = 2.6$ and $3.1$, respectively, for comparison with the particle spectra. 
    The magenta dashed line shows the asymptotic limit of $\alpha = 2.23$ for ultra-relativistic shocks.
    }
    \label{fig:overall_spectra}
\end{figure}

Now we can compare the effect of the different injection nozzles on the particle population in the different jets they generate.
A first point of view is to study the full spectrum of the jets from the different nozzles.
We thus plot the combined 
spectrum of all the particles in the domain. 
This is done by adding the contributions of each of the particles to an overall spectrum.
In Fig.~\ref{fig:overall_spectra} we compare the result for all three nozzles.

For comparison, we also show the initial spectrum (black dashed line) on the same plot. 
This is normalized by taking the initial spectrum of one particle and multiplying it with the total number of 
particles at $t=50$. 
Hence, we have a one to one comparison between the initial and final spectrum for the different nozzles. 

Essentially, we see that compared to the initially injected spectrum, the spectrum for all three nozzles has a significantly larger contribution 
from higher energies. 
Some particles are accelerated to such high energies going even above the initially defined $\gamma_{\rm max} = 10^8$. 
The shape of the spectrum is represented by a broken power-law for the three nozzles, featuring a major peak at $\gamma \simeq 100$. 
This peak is a relic from the initial $\gamma_{\rm min} = 100$. 

For all three nozzles, the spectrum falls for energies below the initial minimum cut-off energy $\gamma_{\rm min}$. 
We emphasize that we do not inject particles at these energies and, thus, the presence of particles at these low energies is a 
result of physical cooling that is implemented in our treatment. 
Interestingly, from the peak at $\gamma = 100$ till $\gamma \simeq 10^{3.5}$, we see the spectrum declining following a power-law 
with a slope of $6$. 
This part of the spectrum is indeed similar to the initial spectrum, which follows a power-law with an index of $6$ as well.
We note that particle are continuously being injected from the jet nozzle.
So, the particles found following the initial spectrum at later times are {\em newly injected particles}, that are
not yet accelerated or cooled down.
Note that the cooling time for those low energy particles is long, and, seemingly, longer than their current age since
injection. 

In general, beyond $\gamma \simeq 10^{3.5}$ the spectrum flattens, following an \textit{ankle} into another power-law. 
The deviation from the injection spectrum is resulting from a number of reasons.
On the the pure increase in number of particles that are accelerated or have cooled.
However, we see that this deviation is also different for the different injection nozzles,
thus, strongly indicating that the overall particle acceleration indeed depends on the jet injection mechanism.

Beyond $\gamma \simeq 10^{3.5}$, the spectrum for setup \textit{prc} flattens the most, following a power-law with index $\alpha \simeq 2.3$
till $\gamma \simeq 10^{4.5}$ where it steepens, forming a \textit{knee}, and another power-law with  $\alpha \simeq 3$. 
The subsequent power law again steepens after passing a second knee at $\gamma \simeq 10^{8.5}$,
again followed by rapid steepening of the spectrum. 

The spectrum for setup \textit{std}, on the other hand, after flattening at $\gamma \simeq 10^{3.5}$, follows a 
power-law with index $\alpha \simeq 2.6$ till $\gamma \simeq 10^{7.5}$, followed by a subsequent decline in the spectra.

Finally, the setup \textit{var} follows a power-law with $\alpha \simeq 3$ between $\gamma \simeq 10^{3.5-6}$. 
Beyond this, the spectrum flattens till a $\gamma \simeq 10^{8}$, and a subsequent rapid steepening. 

Of the three injection nozzles we consider in this study, the precession setup \textit{prc} is clearly the most efficient in 
accelerating particles, with $\gamma$ going up to values even beyond $\sim 10^9$, corresponding to a physical energy reaching 
few hundreds of TeV.
This is followed by he steady injection \textit{std}, while the variable jet setup \textit{var} seems the least efficient of the three
(apart from a narrow range of energies around $\gamma \simeq 10^8$, where it is more efficient than the \textit{std} setup).

After all, and after having in particular compared the energy evolution in various channels for the dynamical evolution in
the previous section,
the interesting question remains about what the energy that is deposited in the {\em particles}. 
We thus now compare the total energy of all the particles in the domain for different injection nozzles setups
for a certain point in time, $t=50$. 

For this purpose, we define the energy of a single (macro-) particle $i$ as 
$E_{{\rm p}, i} = \int_{\gamma_{{\rm min},i}}^{\gamma_{{\rm max},i}} \mathcal{N}(\gamma_i) \gamma_i d\gamma_i$, 
where $\gamma_{{\rm max},i}$ and  $\gamma_{{\rm min},i}$ denote the upper and lower energy cutoffs for that particle, respectively,
and $\gamma_i$ is the Lorentz factor of the electrons of the particle $i$. 
Then, the total energy of all the particles
is $E_{\rm p} = \Sigma_i E_{{\rm p}, i}$.
We find that for the precessing nozzle the particle energy is largest, $E_{\rm p} \simeq 5.9 \times 10^6$ (in code units), followed 
by the variable nozzle with $E_{\rm p} \simeq 5.5 \times 10^6$, and the steady nozzle with $E_{\rm p} \simeq 4.1 \times 10^5$. 

Note at this point that the total particle energy is not only the energy gained by the particles through shocks 
but also contains a portion of the injected particle energy which the particle has to start with. 
In addition, after the particle is injected, the high-energy end of its spectrum decays rapidly as the cooling time for
higher energies is very small as compared to the dynamical time-scales. 
Cooling, of course, also happens to the particles that become shock accelerated, thus a part of the gained energy is lost as well.

Overall, the injected energy of the particles, as well as the gained energy by shocks, can be quickly lost as a result of cooling, 
except at the low-energy regime that have a cooling time larger than the age of the jet (i.e. $t=50$).
This can be seen e.g. in the total particle energy spectra shown in Figure~\ref{fig:overall_spectra},
which are dominated by the injected energy spectrum for energies lower than $\gamma \simeq 10^{3}$. 

Hence, to quantify the level of total energy gained by the particles, we define an {\em effective} total particle energy $E_{\rm p}^{\rm ef}$ 
similar to the total particle energy $E_{\rm p}$, but taking into account only the energy contribution above a threshold of $\gamma = 10^3$. 
Thus, we neglect the injected particle energy and only consider the contribution from the energy regime where the particles were re-energized 
by shocks. 
With that find that the {\em prc} nozzle simulation leads to the highest effective particle energy $E_{\rm p}^{\rm ef} \simeq 8.93 \times 10^3$,
while the nozzles {\rm var} and {\rm std} lead to effective particle energies $E_{\rm p}^{\rm ef} \simeq 0.95 \times 10^3$ and $1.44 \times 10^3$, 
respectively.

Note that these values are only snapshots in time, and a thorough energetics of the particle energy must include considerations
of cooling and re-acceleration along the particle path, involving some time integration of the relevant processes.
We will come back to this issue in our follow-up paper.

\subsection{Spectra of Jet Components}
Astrophysical jets appear as structured beams of material, however, we don't know yet how these structures
form.
We don't know either whether the observed jet sub-structures indicate some radiation pattern or actual material
substructure.
Sub-structure observed in motion could thus be pattern motion or material motion.
One way to disentangle the structure formation process is to study the emission processes of the material.
Here, we compare the particle spectra of the different outflow components. 

\begin{figure}[t]
    \centering
    \includegraphics[width = 1.0\linewidth]{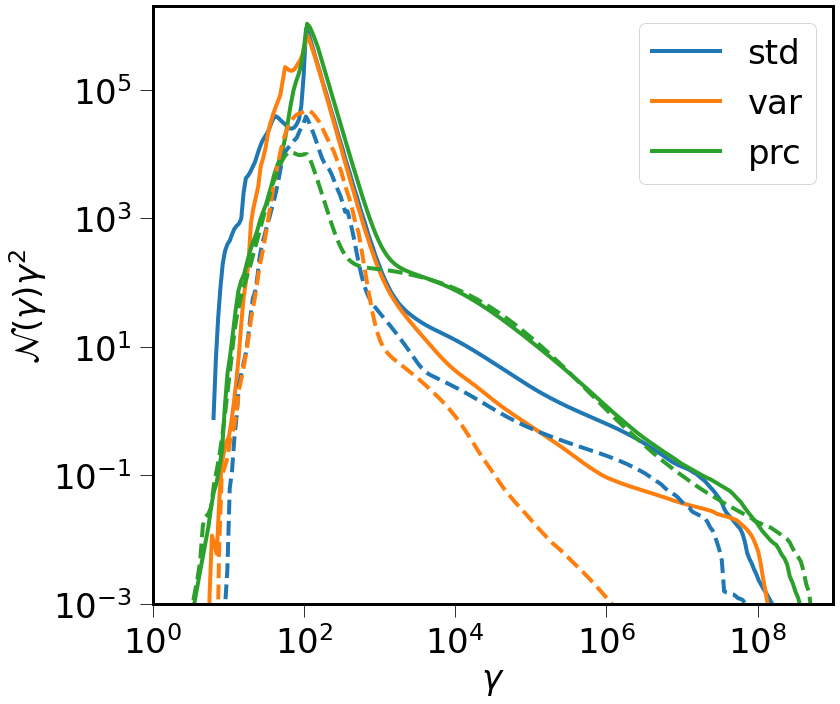}
    \caption{Energy spectra considering all the particles in the jet structure (solid line) and the surrounding entrained material (dashed line) for
     the simulation runs \textit{std} (blue), \textit{var} (orange) and \textit{prc} (green) at a time $t =50$ (in code units). 
     Here, the particles in the jet and the entrained material are defined by having a tracer value $>0.8$ and $<0.8$, respectively.}
    \label{fig:jet_cocoon_spectra}
\end{figure}

In order to investigate the role of the different jet components for the acceleration of particles, 
we have therefore studied the spectrum from these regions. 

Along the length of the jet, we are interested in the spectrum near the jet base, the jet head, and at some intermediary 
region in the jet. 
Note that comparing spatially different areas \textit{along} the jet also corresponds to investigate a time evolution 
of these features, as, due to the jet propagation, these areas have evolved for different periods of time after injection.
Of particular interest is  steady {\rev internal} shock feature we have observed in simulation run \textit{std}, where we can as well
study its effect on the particle acceleration and spectrum. 

In addition to substructures along the jet, we want to compare substructures across the jet.
We thus distinguish between the particles in the jet and in the entrained material (a mix of the ambient and jet material), and investigate particle acceleration in these areas.
In this regard, we define the \textit{jet} and \textit{entrained} material as the regions where the value of tracer is $\geq 0.8$ or $< 0.8$, 
respectively. 
We show the spectral energy distribution for the particles in the jet and entrained material for all three nozzles in 
Figure~\ref{fig:jet_cocoon_spectra}. 

For the steady injection nozzle, the jet is dominant in accelerating particles as compared to the entrained material at all particle energies. 
This is also true for the nozzle with variable injection. 
However, we find that the material entrained by the steady nozzle setup \textit{std} is much more efficient than that in the \textit{var} nozzle setup in order to accelerate 
particles to energies beyond $\gamma \simeq 10^4$. 

In contrast, the entrained material in the precessing jet is a much more efficient accelerator. 
Here the spectra from particles located in the entrained material have a contribution comparable to those located in the jet for the energy range $\gamma = 10^{3.5}$ to $10^6$. 
At the highest particle energies, $\gamma \geq 8.5$, for the setup with the precessing nozzle the acceleration in the entrained material surpasses even that in the
corresponding jet. 
This is ultimately a consequence of the dynamics of the precessing nozzle. 
The gas in the entrained material in this setup is being constantly hit and perturbed by the jet material, generating many shocks, and subsequent enhanced shock acceleration in the entrained material.

To study the effect of {\rev strong steady internal shock} feature in the steady nozzle setup {\em std}, we analyze the spectra from the area blocks of region above, below and 
enclosing the steady {\rev internal} shock. 
These blocks are essentially cylindrical slices with radius spanning the whole domain\footnote{Although the size of the blocks span the whole domain,
it should be noted that the particles are concentrated just in the jet and the cocoon and there are no particles in the ambient medium. }
and a height of $\Delta z =1$, centered at $z = 2$ (below the steady  {\rev internal}
shock region, at the jet base), $z=14$ (in the steady {\rev internal} shock region), and $z=20$ (above the steady {\rev internal} shock region). 
Hence, the {\em average} spectrum from a block is the sum of spectra from each of the particles inside the block normalized to the number of particles in the block. 
Additionally, we also analyze the spectrum from the {\em jet head} which we define as the region where $z > 32$.

\begin{figure}[t]
    \centering
    \includegraphics[width = 0.9\linewidth]{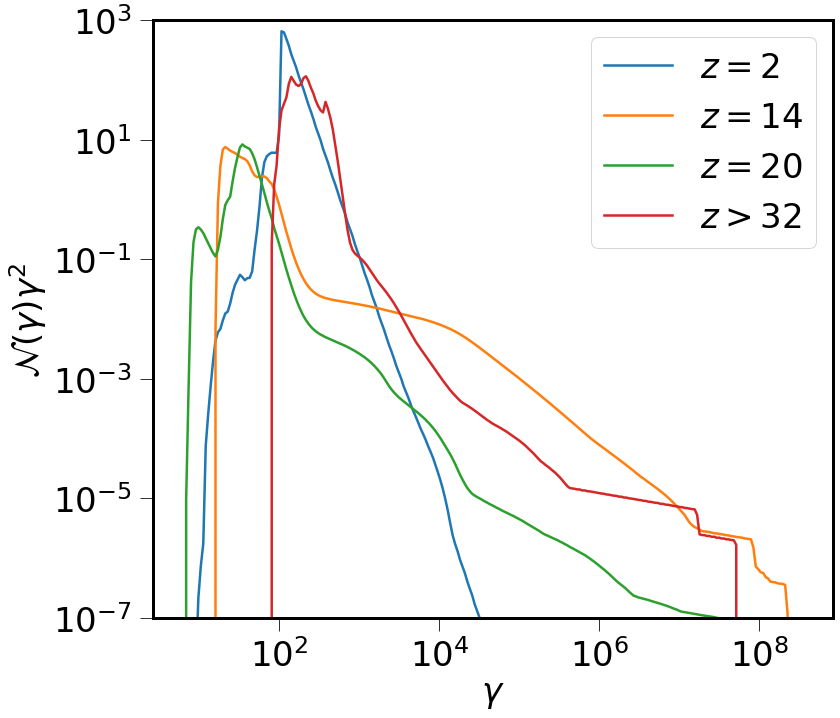}
    \caption{Average energy spectra of all the particles in the cylindrical slices of height $1$ (in code units) centered at $z = 6$ (below the steady {\rev internal} shock region),
     $z=14$ (in the steady {\rev internal}  shock region), and $z=20$ (above the steady {\rev internal} shock region). }
    \label{fig:ss_spectra}
\end{figure}

We expect a stronger acceleration for particles in the block located at $z=14$ due to the presence of the strong shock around that location. 
We expect the same for the particles in the jet head, owing to the presence of a termination shock there.
This is precisely what we see in Figure~\ref{fig:ss_spectra} where we show the spectra from these four blocks. 
We clearly see a larger contribution to the spectrum for energies beyond $\gamma \simeq 10^{4}$ from the blocks at $z=14$ and $z>32$ as compared to the
two other blocks. 

Interestingly, the steady {\rev internal} shock is capable of accelerating particles to energies higher than $\gamma \simeq 10^8$ (corresponding to electrons with energies of few tens of TeV),
surpassing the efficiency of the termination shock as well.
Since this shock at {\rev $z=14.8$} is (i) steady and (ii) filled with high energetic particles, we expect to see a stationary {\em knot}-like feature in the 
emission map (which we will explore in a subsequent paper) at this location. 

As discussed above, comparing spatially different areas along the jet also corresponds to study the time-evolution of the jet spectrum. 
This is also evident in Figure~\ref{fig:ss_spectra}. 
At $z=2$, we are in the close vicinity of the jet base from where the {\em new} particles are being continuously injected. 
As a result, the spectrum at that location closely resemble the initially injected spectrum with a peak at $\gamma \simeq 100$,
a consequence of the initial minimum cutoff $\gamma = 100$. 
Beyond this, the spectrum follows a power-law with slope $\simeq 6$, same as the initially injected power-law index $\alpha = 6$. 
As we move downstream along the jet, the particles enter the steady {\rev internal} shock and are accelerated to high energies, as discussed. 
Further downstream, at $z=20$, the particles have cooled down after leaving the strong shock to lower energies, decreasing both the upper and 
lower energy cutoffs. 
At last, particles meet the termination shock and are accelerated again to high energies.

\section{Conclusions}
\label{sec:summary}
We have applied the PLUTO code \citep{mignone2007pluto} to perform 3D relativistic MHD simulations in combination with a hybrid
module \citep{vaidya2018} that is capable of following Lagrangian macro-particles with the fluid.
The injected Lagrangian macro-particles constitute ensembles of electrons that are close in space.
Their initially defined energy distribution changes as they evolve in time, considering diffusive shock 
acceleration and cooling due to synchrotron radiation, inverse Compton processes and adiabatic expansion of the jet. 

We investigate how physically different jet injection nozzles affect the particle spectrum 
that is generated as a consequence of the underlying jet dynamics.
We have considered three different nozzle setups, a standard steady jet injection into an
unstratified ambient gas, a time-dependent jet injection, and a jet nozzle undergoing precession.
We adopt jet injection profiles applying an equilibrium solution from \citet{bodo2019}. 
This is essential in order to avoid artificial instabilities that mimic turbulence and, hence, lead
to unphysical particle acceleration. 
We adopt a Taub-Matthews equation of state, as we consider a relativistic gas embedded in a non-relativistic ambient medium.

Considering our resolution study, we apply a resolution of 25 cells per jet radius,
more than ever applied in 3D MHD jet simulations.
This resolution is needed for both, resolving dynamically interesting features, as well as capturing the structure and 
strength of the shocks that are involved in particle acceleration.

While our simulations are run in code units and are applicable to any relativistic jet source, from the prescribed injection
Lorentz factor of 10, the corresponding astrophysical scaling of our simulations would be that of a pc-scale AGN jet, with a 
size of our simulation domain of $10 \times 10 \times 20$ pc.
Naturally, micro-quasars with a similar initial speed can also be described, however, the physical magnetic field strength 
we have to assume for particle cooling, is chosen for AGN-like conditions.

Our results are as follows. 

(i) For all nozzle geometries investigated the interaction of injected jet material with the ambient gas and with
the previously injected gas leads to the formation of strong shocks.
Essentially, the location and the properties of these shocks vary for the different nozzles.

(ii) The steadily injected jet forms a well known jet-cocoon structure. 
In this simulation we find a special shock feature -- a strong and stationary shock resulting
from head on collision of jet and a back-flow along the jet axis {\em inside} the jet.
The location and structure of this feature converges only for sufficiently high numerical resolution, 
i.e.~25 cells per jet radius.

(iii) The jet with variable injection evolves into a narrower cocoon, resulting in an overall structure more collimated
than the steady jet.
This jet forms multiple bow shocks as a result of various episodes of time-dependent activity, leading 
to a skeletal appearance. 

(iv) The dynamics of the precessing jet is more turbulent as compared to the other nozzles we consider. 
As the jet axis change over time, multiple shocks are observed at the positions where the 
(multiple) jet heads interact with the ambient medium and with the jet.

(v) We quantify the level of turbulence mainly by the fluctuations of the energy density.
We compare the energy budget involved in turbulent motions with the mean fluid motion and other energy
channels. We find that about 1\% of the bulk energies is turned into the turbulent energy channels.

(vi) On encountering a shock, the injected Lagrangian macro-particles change their spectral energy distribution
depending on the shock compression ratio they experience. 
We find that the variable jet injection nozzle produces the strongest shocks (with more of the highest compression ratios), 
followed by the precessing jet and the steady jet. 
On the other hand, the total number of shocks is largest for the precessing jet, followed by the steady and the variable jet. 
The overall particle acceleration depends on the interplay between the available shock strength and the number of shocks 
they experience. 

(vii) The macro-particles cool down due to synchrotron, inverse Compton and expansion losses. 
While particle acceleration happens instantaneously when a macro-particle crosses a shock, cooling depends
on magnetic field strength and particle energy.
For the chosen magnetic field strength, the cooling time for the high energy tail of the particle energy spectrum
is well below the dynamical time scale of our jet simulations.

(viii) The total energy spectral distribution induced by the different nozzles shows a significant increase in the 
number of high-energy particles over time.
The precessing jet leads to the formation of comparatively more shocks that are most efficient in accelerating 
particles, followed by the steady jet, while the shocks in the variable jet are the least efficient of the three.
In the precessing jet electron are accelerated up to a few hundreds TeV with a power-law index
of $\alpha \simeq 2.3$ to $3.1$.
For the steady and the variable jets electron energies are up to a few tens of TeV, with spectra of similar
but somewhat steeper slopes.
From this we find that particle acceleration is governed more by the total number of shocks available than
by the strength of the shocks.

(ix) We have compared various regions of the jet concerning their particle energy distribution.
Compared to the jet-entrained ambient material the jet plays a dominant role for 
the particle acceleration for the steady nozzle and the variable injection nozzle. 
For the variable nozzle, the particle acceleration in the entrained structure is highly suppressed for higher electron energies.
For the precessing jet, entrained material and jet are comparable in particle acceleration efficiency, except at the very high-energy 
end, where the entrained material is even more dominant than the jet for acceleration. 
Here, particles in the entrained material are being re-energized through shocks continuously -- a result of the highly turbulent jet dynamics. 
Hence, we expect precessing AGN jets to be the best sites for highly efficient particle acceleration. 

(x) The total energy of accelerated particles in the precessing and variable jet is higher than the steady jet as a result
of more shocks available due to the turbulent motion induced by the time-dependent injection. 
The energy the particles have gained till the end of the simulation can be up to 5\% of the turbulent energy, i.e. below $\simeq 1\%$ of the
injected jet energy. 

(xi) We have studied the effect of the strong, stationary shock feature that we have discovered, on the particle spectrum. 
Like the jet termination shock, this strong stationary shock is a particular site where electrons are accelerated to very high energies of few tens of TeV. 
As these highly energized particles cool rapidly, we expect that this dynamical feature can be observed as a stationary knot of enhanced intensity in emission. 
The similarity to the famous HST-1 knot seems intriguing. 

To summarize we have demonstrated and quantified how different jet nozzles convert the injected jet power to different levels of turbulence,
which in turn leads to the generation of shocks that can accelerate electrons to energies up to tens and even hundreds of TeV. 
The resulting particle energies are about 10\% of the turbulent energy and below 1\% of the available jet energy. 

In a subsequent paper, we will apply our results presenting synthetic emission maps of these jets.
This will require a specific choice of the physical scaling for densities and pressure, and can be applied for different
relativistic jet sources, such as AGN or micro quasars, as well as for different viewing angles.

\acknowledgements
This project is financed through grant DFG-FOR5195 of the German Science Foundation DFG.
B.V. acknowledges funding by the Max Planck Partner Group located at IIT Indore. 
R.D. acknowledges travel funds by the International Max Planck Research School for Astronomy \& Cosmic Physics at the University of Heidelberg.
We thank Andrea Mignone for providing and sustaining the PLUTO code.
We acknowledge enlightening discussions with Karl Mannheim and John Kirk.
All simulations were performed at the MPCDF computing center of the Max Planck Society
in Garching (utilizing the compute clusters Vera, Raven \& Cobra).
We are grateful to an unknown referee for a detailed and helpful report that considerably improved
the clarity of the paper.


\bibliography{main}{}

\begin{thebibliography}{}
\expandafter\ifx\csname natexlab\endcsname\relax\def\natexlab#1{#1}\fi
\providecommand{\url}[1]{\href{#1}{#1}}
\providecommand{\dodoi}[1]{doi:~\href{http://doi.org/#1}{\nolinkurl{#1}}}
\providecommand{\doeprint}[1]{\href{http://ascl.net/#1}{\nolinkurl{http://ascl.net/#1}}}
\providecommand{\doarXiv}[1]{\href{https://arxiv.org/abs/#1}{\nolinkurl{https://arxiv.org/abs/#1}}}

\bibitem[{{Aab} {et~al.}(2018){Aab}, {Abreu}, {Aglietta}, {Albuquerque},
  {Allekotte}, {Almela}, {Alvarez Castillo}, {Alvarez-Mu{\~n}iz}, {Anastasi},
  {Anchordoqui}, {Andrada}, {Andringa}, {Aramo}, {Arsene}, {Asorey}, {Assis},
  {Avila}, {Badescu}, {Balaceanu}, {Barbato}, {Barreira Luz}, {Beatty},
  {Becker}, {Bellido}, {Berat}, {Bertaina}, {Bertou}, {Biermann}, {Biteau},
  {Blaess}, {Blanco}, {Blazek}, {Bleve}, {Boh{\'a}{\v{c}}ov{\'a}}, {Bonifazi},
  {Borodai}, {Botti}, {Brack}, {Brancus}, {Bretz}, {Bridgeman}, {Briechle},
  {Buchholz}, {Bueno}, {Buitink}, {Buscemi}, {Caballero-Mora}, {Caccianiga},
  {Cancio}, {Canfora}, {Caruso}, {Castellina}, {Catalani}, {Cataldi}, {Cazon},
  {Chavez}, {Chinellato}, {Chudoba}, {Clay}, {Cobos Cerutti}, {Colalillo},
  {Coleman}, {Collica}, {Coluccia}, {Concei{\c{c}}{\~a}o}, {Consolati},
  {Contreras}, {Cooper}, {Coutu}, {Covault}, {Cronin}, {D'Amico}, {Daniel},
  {Dasso}, {Daumiller}, {Dawson}, {de Almeida}, {de Jong}, {De Mauro}, {de
  Mello Neto}, {De Mitri}, {de Oliveira}, {de Souza}, {Debatin}, {Deligny},
  {D{\'\i}az Castro}, {Diogo}, {Dobrigkeit}, {D'Olivo}, {Dorosti}, {dos Anjos},
  {Dova}, {Dundovic}, {Ebr}, {Engel}, {Erdmann}, {Erfani}, {Escobar},
  {Espadanal}, {Etchegoyen}, {Falcke}, {Farmer}, {Farrar}, {Fauth}, {Fazzini},
  {Fenu}, {Fick}, {Figueira}, {Filip{\v{c}}i{\v{c}}}, {Freire}, {Fujii},
  {Fuster}, {Ga{\"\i}or}, {Garc{\'\i}a}, {Gat{\'e}}, {Gemmeke},
  {Gherghel-Lascu}, {Ghia}, {Giaccari}, {Giammarchi}, {Giller}, {G{\l}as},
  {Glaser}, {Golup}, {G{\'o}mez Berisso}, {G{\'o}mez Vitale}, {Gonz{\'a}lez},
  {Gorgi}, {Grillo}, {Grubb}, {Guarino}, {Guedes}, {Halliday}, {Hampel},
  {Hansen}, {Harari}, {Harrison}, {Haungs}, {Hebbeker}, {Heck}, {Heimann},
  {Herve}, {Hill}, {Hojvat}, {Holt}, {Homola}, {H{\"o}randel}, {Horvath},
  {Hrabovsk{\'y}}, {Huege}, {Hulsman}, {Insolia}, {Isar}, {Jandt}, {Johnsen},
  {Josebachuili}, {Jurysek}, {K{\"a}{\"a}p{\"a}}, {Kambeitz}, {Kampert},
  {Keilhauer}, {Kemmerich}, {Kemp}, {Kemp}, {Kieckhafer}, {Klages}, {Kleifges},
  {Kleinfeller}, {Krause}, {Krohm}, {Kuempel}, {Kukec Mezek}, {Kunka}, {Kuotb
  Awad}, {Lago}, {LaHurd}, {Lang}, {Lauscher}, {Legumina}, {Leigui de
  Oliveira}, {Letessier-Selvon}, {Lhenry-Yvon}, {Link}, {Lo Presti}, {Lopes},
  {L{\'o}pez}, {L{\'o}pez Casado}, {Lorek}, {Luce}, {Lucero}, {Malacari},
  {Mallamaci}, {Mandat}, {Mantsch}, {Mariazzi}, {Mari{\c{s}}}, {Marsella},
  {Martello}, {Martinez}, {Mart{\'\i}nez Bravo}, {Mas{\'\i}as Meza}, {Mathes},
  {Mathys}, {Matthews}, {Matthiae}, {Mayotte}, {Mazur}, {Medina},
  {Medina-Tanco}, {Melo}, {Menshikov}, {Merenda}, {Michal}, {Micheletti},
  {Middendorf}, {Miramonti}, {Mitrica}, {Mockler}, {Mollerach}, {Montanet},
  {Morello}, {Morlino}, {Mostaf{\'a}}, {M{\"u}ller}, {M{\"u}ller}, {Muller},
  {M{\"u}ller}, {Mussa}, {Naranjo}, {Nellen}, {Nguyen}, {Niculescu-Oglinzanu},
  {Niechciol}, {Niemietz}, {Niggemann}, {Nitz}, {Nosek}, {Novotny},
  {No{\v{z}}ka}, {N{\'u}{\~n}ez}, {Oikonomou}, {Olinto}, {Palatka}, {Pallotta},
  {Papenbreer}, {Parente}, {Parra}, {Paul}, {Pech}, {Pedreira}, {P{\c{e}}kala},
  {Pelayo}, {Pe{\~n}a-Rodriguez}, {Pereira}, {Perlin}, {Perrone}, {Peters},
  {Petrera}, {Phuntsok}, {Pierog}, {Pimenta}, {Pirronello}, {Platino}, {Plum},
  {Poh}, {Porowski}, {Prado}, {Privitera}, {Prouza}, {Quel}, {Querchfeld},
  {Quinn}, {Ramos-Pollan}, {Rautenberg}, {Ravignani}, {Ridky}, {Riehn},
  {Risse}, {Ristori}, {Rizi}, {Rodrigues de Carvalho}, {Rodriguez Fernandez},
  {Rodriguez Rojo}, {Roncoroni}, {Roth}, {Roulet}, {Rovero}, {Ruehl}, {Saffi},
  {Saftoiu}, {Salamida}, {Salazar}, {Saleh}, {Salina}, {S{\'a}nchez},
  {Sanchez-Lucas}, {Santos}, {Santos}, {Sarazin}, {Sarmento}, {Sarmiento-Cano},
  {Sato}, {Schauer}, {Scherini}, {Schieler}, {Schimp}, {Schmidt}, {Scholten},
  {Schov{\'a}nek}, {Schr{\"o}der}, {Schr{\"o}der}, {Schulz}, {Schumacher},
  {Sciutto}, {Segreto}, {Shadkam}, {Shellard}, {Sigl}, {Silli},
  {{\v{S}}m{\'\i}da}, {Snow}, {Sommers}, {Sonntag}, {Soriano}, {Squartini},
  {Stanca}, {Stani{\v{c}}}, {Stasielak}, {Stassi}, {Stolpovskiy}, {Strafella},
  {Streich}, {Suarez}, {Suarez Dur{\'a}n}, {Sudholz}, {Suomij{\"a}rvi},
  {Supanitsky}, {{\v{S}}up{\'\i}k}, {Swain}, {Szadkowski}, {Taboada},
  {Taborda}, {Theodoro}, {Timmermans}, {Todero Peixoto}, {Tomankova},
  {Tom{\'e}}, {Torralba Elipe}, {Travnicek}, {Trini}, {Ulrich}, {Unger},
  {Urban}, {Vald{\'e}s Galicia}, {Vali{\~n}o}, {Valore}, {van Aar}, {van
  Bodegom}, {van den Berg}, {van Vliet}, {Varela}, {Vargas C{\'a}rdenas},
  {V{\'a}zquez}, {Veberi{\v{c}}}, {Ventura}, {Vergara Quispe}, {Verzi},
  {Vicha}, {Villase{\~n}or}, {Vorobiov}, {Wahlberg}, {Wainberg}, {Walz},
  {Watson}, {Weber}, {Weindl}, {Wiede{\'n}ski}, {Wiencke}, {Wilczy{\'n}ski},
  {Wirtz}, {Wittkowski}, {Wundheiler}, {Yang}, {Yushkov}, {Zas}, {Zavrtanik},
  {Zavrtanik}, {Zepeda}, {Zimmermann}, {Ziolkowski}, {Zong}, {Zuccarello}, \&
  {Pierre Auger Collaboration}}]{aab2018}
{Aab}, A., {Abreu}, P., {Aglietta}, M., {et~al.} 2018, \apjl, 853, L29,
  \dodoi{10.3847/2041-8213/aaa66d}

\bibitem[{{Appl} \& {Camenzind}(1988)}]{appl1988}
{Appl}, S., \& {Camenzind}, M. 1988, \aap, 206, 258

\bibitem[{{Araudo} {et~al.}(2021){Araudo}, {Padovani}, \&
  {Marcowith}}]{araudo2021}
{Araudo}, A.~T., {Padovani}, M., \& {Marcowith}, A. 2021, \mnras, 504, 2405,
  \dodoi{10.1093/mnras/stab635}

\bibitem[{{Bell}(1978{\natexlab{a}})}]{bell1978a}
{Bell}, A.~R. 1978{\natexlab{a}}, \mnras, 182, 147,
  \dodoi{10.1093/mnras/182.2.147}

\bibitem[{{Bell}(1978{\natexlab{b}})}]{bell1978b}
---. 1978{\natexlab{b}}, \mnras, 182, 443, \dodoi{10.1093/mnras/182.3.443}

\bibitem[{{Bell} {et~al.}(2018){Bell}, {Araudo}, {Matthews}, \&
  {Blundell}}]{bell2018}
{Bell}, A.~R., {Araudo}, A.~T., {Matthews}, J.~H., \& {Blundell}, K.~M. 2018,
  \mnras, 473, 2364, \dodoi{10.1093/mnras/stx2485}

\bibitem[{{Biretta} {et~al.}(1999){Biretta}, {Sparks}, \&
  {Macchetto}}]{biretta1999}
{Biretta}, J.~A., {Sparks}, W.~B., \& {Macchetto}, F. 1999, \apj, 520, 621,
  \dodoi{10.1086/307499}

\bibitem[{{Blandford} \& {McKee}(1976)}]{blandord1976}
{Blandford}, R.~D., \& {McKee}, C.~F. 1976, Physics of Fluids, 19, 1130,
  \dodoi{10.1063/1.861619}

\bibitem[{{Blandford} \& {Ostriker}(1978)}]{blandford1978}
{Blandford}, R.~D., \& {Ostriker}, J.~P. 1978, \apjl, 221, L29,
  \dodoi{10.1086/182658}

\bibitem[{{Blandford} \& {Payne}(1982)}]{BP1982}
{Blandford}, R.~D., \& {Payne}, D.~G. 1982, \mnras, 199, 883,
  \dodoi{10.1093/mnras/199.4.883}

\bibitem[{{Blandford} \& {Znajek}(1977)}]{BZ1977}
{Blandford}, R.~D., \& {Znajek}, R.~L. 1977, \mnras, 179, 433,
  \dodoi{10.1093/mnras/179.3.433}

\bibitem[{{Bodo} {et~al.}(2019){Bodo}, {Mamatsashvili}, {Rossi}, \&
  {Mignone}}]{bodo2019}
{Bodo}, G., {Mamatsashvili}, G., {Rossi}, P., \& {Mignone}, A. 2019, \mnras,
  485, 2909, \dodoi{10.1093/mnras/stz591}

\bibitem[{{Britzen} {et~al.}(2018){Britzen}, {Fendt}, {Witzel}, {Qian},
  {Pashchenko}, {Kurtanidze}, {Zajacek}, {Martinez}, {Karas}, {Aller}, {Aller},
  {Eckart}, {Nilsson}, {Ar{\'e}valo}, {Cuadra}, {Subroweit}, \&
  {Witzel}}]{britzen2018}
{Britzen}, S., {Fendt}, C., {Witzel}, G., {et~al.} 2018, \mnras, 478, 3199,
  \dodoi{10.1093/mnras/sty1026}

\bibitem[{{Britzen} {et~al.}(2019){Britzen}, {Fendt}, {B{\"o}ttcher},
  {Zaja{\v{c}}ek}, {Jaron}, {Pashchenko}, {Araudo}, {Karas}, \&
  {Kurtanidze}}]{britzen2019a}
{Britzen}, S., {Fendt}, C., {B{\"o}ttcher}, M., {et~al.} 2019, \aap, 630, A103,
  \dodoi{10.1051/0004-6361/201935422}

\bibitem[{{Brunetti} {et~al.}(2003){Brunetti}, {Mack}, {Prieto}, \&
  {Varano}}]{brunetti2003}
{Brunetti}, G., {Mack}, K.~H., {Prieto}, M.~A., \& {Varano}, S. 2003, \mnras,
  345, L40, \dodoi{10.1046/j.1365-8711.2003.07185.x}

\bibitem[{{Carilli} {et~al.}(1999){Carilli}, {Kurk}, {van der Werf}, {Perley},
  \& {Miley}}]{carilli1999}
{Carilli}, C.~L., {Kurk}, J.~D., {van der Werf}, P.~P., {Perley}, R.~A., \&
  {Miley}, G.~K. 1999, \aj, 118, 2581, \dodoi{10.1086/301137}

\bibitem[{{Casse} \& {Keppens}(2002)}]{keppens2002}
{Casse}, F., \& {Keppens}, R. 2002, \apj, 581, 988, \dodoi{10.1086/344340}

\bibitem[{{Del Zanna} {et~al.}(2003){Del Zanna}, {Bucciantini}, \&
  {Londrillo}}]{delzanna2003}
{Del Zanna}, L., {Bucciantini}, N., \& {Londrillo}, P. 2003, \aap, 400, 397,
  \dodoi{10.1051/0004-6361:20021641}

\bibitem[{{Dihingia} {et~al.}(2021){Dihingia}, {Vaidya}, \&
  {Fendt}}]{dihingia2021}
{Dihingia}, I.~K., {Vaidya}, B., \& {Fendt}, C. 2021, \mnras, 505, 3596,
  \dodoi{10.1093/mnras/stab1512}

\bibitem[{{Ehlert} {et~al.}(2018){Ehlert}, {Weinberger}, {Pfrommer}, {Pakmor},
  \& {Springel}}]{ehlert2018}
{Ehlert}, K., {Weinberger}, R., {Pfrommer}, C., {Pakmor}, R., \& {Springel}, V.
  2018, \mnras, 481, 2878, \dodoi{10.1093/mnras/sty2397}

\bibitem[{{Fanaroff} \& {Riley}(1974)}]{FR1974}
{Fanaroff}, B.~L., \& {Riley}, J.~M. 1974, \mnras, 167, 31P,
  \dodoi{10.1093/mnras/167.1.31P}

\bibitem[{{Fermi}(1949)}]{fermi1949}
{Fermi}, E. 1949, Physical Review, 75, 1169, \dodoi{10.1103/PhysRev.75.1169}

\bibitem[{{Ferreira} \& {Pelletier}(1995)}]{ferreira1995}
{Ferreira}, J., \& {Pelletier}, G. 1995, \aap, 295, 807

\bibitem[{{Giannios} {et~al.}(2009){Giannios}, {Uzdensky}, \&
  {Begelman}}]{giannios2009}
{Giannios}, D., {Uzdensky}, D.~A., \& {Begelman}, M.~C. 2009, \mnras, 395, L29,
  \dodoi{10.1111/j.1745-3933.2009.00635.x}

\bibitem[{{Giri} {et~al.}(2022){Giri}, {Dubey}, {Rubinur}, {Vaidya}, \&
  {Kharb}}]{giri2022}
{Giri}, G., {Dubey}, R.~P., {Rubinur}, K., {Vaidya}, B., \& {Kharb}, P. 2022,
  \mnras, 514, 5625, \dodoi{10.1093/mnras/stac1628}

\bibitem[{{Guess}(1960)}]{guess1960}
{Guess}, A.~W. 1960, Physics of Fluids, 3, 697, \dodoi{10.1063/1.1706113}

\bibitem[{{Hargrave} \& {Ryle}(1974)}]{hargrave1974}
{Hargrave}, P.~J., \& {Ryle}, M. 1974, \mnras, 166, 305,
  \dodoi{10.1093/mnras/166.2.305}

\bibitem[{{Harris} {et~al.}(2006){Harris}, {Cheung}, {Biretta}, {Sparks},
  {Junor}, {Perlman}, \& {Wilson}}]{harris2006}
{Harris}, D.~E., {Cheung}, C.~C., {Biretta}, J.~A., {et~al.} 2006, \apj, 640,
  211, \dodoi{10.1086/500081}

\bibitem[{{Hawley} {et~al.}(2015){Hawley}, {Fendt}, {Hardcastle}, {Nokhrina},
  \& {Tchekhovskoy}}]{hawley2015}
{Hawley}, J.~F., {Fendt}, C., {Hardcastle}, M., {Nokhrina}, E., \&
  {Tchekhovskoy}, A. 2015, \ssr, 191, 441, \dodoi{10.1007/s11214-015-0174-7}

\bibitem[{{Heavens} \& {Meisenheimer}(1987)}]{heavens1987}
{Heavens}, A.~F., \& {Meisenheimer}, K. 1987, \mnras, 225, 335,
  \dodoi{10.1093/mnras/225.2.335}

\bibitem[{{Hervet} {et~al.}(2016){Hervet}, {Boisson}, \& {Sol}}]{hervet2016}
{Hervet}, O., {Boisson}, C., \& {Sol}, H. 2016, \aap, 592, A22,
  \dodoi{10.1051/0004-6361/201628117}

\bibitem[{{Huang} {et~al.}(2023){Huang}, {Reville}, {Kirk}, \&
  {Giacinti}}]{huang2023}
{Huang}, Z.-Q., {Reville}, B., {Kirk}, J.~G., \& {Giacinti}, G. 2023, arXiv
  e-prints, arXiv:2304.08132, \dodoi{10.48550/arXiv.2304.08132}

\bibitem[{{Keppens} {et~al.}(2008){Keppens}, {Meliani}, {van der Holst}, \&
  {Casse}}]{Keppens2008}
{Keppens}, R., {Meliani}, Z., {van der Holst}, B., \& {Casse}, F. 2008, \aap,
  486, 663, \dodoi{10.1051/0004-6361:20079174}

\bibitem[{{Kharb} {et~al.}(2017){Kharb}, {Lal}, \& {Merritt}}]{kharb2017}
{Kharb}, P., {Lal}, D.~V., \& {Merritt}, D. 2017, Nature Astronomy, 1, 727,
  \dodoi{10.1038/s41550-017-0256-4}

\bibitem[{{Kirk} \& {Duffy}(1999)}]{kirk1999}
{Kirk}, J.~G., \& {Duffy}, P. 1999, Journal of Physics G Nuclear Physics, 25,
  R163, \dodoi{10.1088/0954-3899/25/8/201}

\bibitem[{{Kirk} {et~al.}(2000){Kirk}, {Guthmann}, {Gallant}, \&
  {Achterberg}}]{kirk2000}
{Kirk}, J.~G., {Guthmann}, A.~W., {Gallant}, Y.~A., \& {Achterberg}, A. 2000,
  \apj, 542, 235, \dodoi{10.1086/309533}

\bibitem[{{Konigl}(1980)}]{koenigl1980}
{Konigl}, A. 1980, Physics of Fluids, 23, 1083, \dodoi{10.1063/1.863110}

\bibitem[{{Krause} {et~al.}(2019){Krause}, {Shabala}, {Hardcastle}, {Bicknell},
  {B{\"o}hringer}, {Chon}, {Nawaz}, {Sarzi}, \& {Wagner}}]{krause2019}
{Krause}, M. G.~H., {Shabala}, S.~S., {Hardcastle}, M.~J., {et~al.} 2019,
  \mnras, 482, 240, \dodoi{10.1093/mnras/sty2558}

\bibitem[{{Krymskii} {et~al.}(1978){Krymskii}, {Elshin}, {Romashchenko},
  {Transkii}, {Gusev}, \& {Altukhov}}]{krymskii1978}
{Krymskii}, G.~F., {Elshin}, V.~K., {Romashchenko}, I.~A., {et~al.} 1978,
  Akademiia Nauk SSSR Izvestiia Seriia Fizicheskaia, 42, 1075

\bibitem[{{Kundu} {et~al.}(2021){Kundu}, {Vaidya}, \& {Mignone}}]{kundu2021}
{Kundu}, S., {Vaidya}, B., \& {Mignone}, A. 2021, \apj, 921, 74,
  \dodoi{10.3847/1538-4357/ac1ba5}

\bibitem[{{Leismann} {et~al.}(2005){Leismann}, {Ant{\'o}n}, {Aloy},
  {M{\"u}ller}, {Mart{\'\i}}, {Miralles}, \& {Ib{\'a}{\~n}ez}}]{leismann2005}
{Leismann}, T., {Ant{\'o}n}, L., {Aloy}, M.~A., {et~al.} 2005, \aap, 436, 503,
  \dodoi{10.1051/0004-6361:20042520}

\bibitem[{{Lichnerowicz}(1976)}]{lichnerowicz1976}
{Lichnerowicz}, A. 1976, Journal of Mathematical Physics, 17, 2135,
  \dodoi{10.1063/1.522857}

\bibitem[{{Lister} {et~al.}(2021){Lister}, {Homan}, {Kellermann}, {Kovalev},
  {Pushkarev}, {Ros}, \& {Savolainen}}]{lister2021}
{Lister}, M.~L., {Homan}, D.~C., {Kellermann}, K.~I., {et~al.} 2021, \apj, 923,
  30, \dodoi{10.3847/1538-4357/ac230f}

\bibitem[{{Mannheim}(1993)}]{mannheim1993}
{Mannheim}, K. 1993, \aap, 269, 67, \dodoi{10.48550/arXiv.astro-ph/9302006}

\bibitem[{{Marcowith} {et~al.}(2020){Marcowith}, {Ferrand}, {Grech}, {Meliani},
  {Plotnikov}, \& {Walder}}]{marcowith2020}
{Marcowith}, A., {Ferrand}, G., {Grech}, M., {et~al.} 2020, Living Reviews in
  Computational Astrophysics, 6, 1, \dodoi{10.1007/s41115-020-0007-6}

\bibitem[{{Marcowith} {et~al.}(2016){Marcowith}, {Bret}, {Bykov}, {Dieckman},
  {O'C Drury}, {Lemb{\`e}ge}, {Lemoine}, {Morlino}, {Murphy}, {Pelletier},
  {Plotnikov}, {Reville}, {Riquelme}, {Sironi}, \& {Stockem
  Novo}}]{marcowith2016}
{Marcowith}, A., {Bret}, A., {Bykov}, A., {et~al.} 2016, Reports on Progress in
  Physics, 79, 046901, \dodoi{10.1088/0034-4885/79/4/046901}

\bibitem[{{Mathews}(1971)}]{mathews1971}
{Mathews}, W.~G. 1971, \apj, 165, 147, \dodoi{10.1086/150883}

\bibitem[{{Matthews} {et~al.}(2018){Matthews}, {Bell}, {Blundell}, \&
  {Araudo}}]{matthews2018}
{Matthews}, J.~H., {Bell}, A.~R., {Blundell}, K.~M., \& {Araudo}, A.~T. 2018,
  \mnras, 479, L76, \dodoi{10.1093/mnrasl/sly099}

\bibitem[{{Matthews} {et~al.}(2019){Matthews}, {Bell}, {Blundell}, \&
  {Araudo}}]{Matthews2019}
---. 2019, \mnras, 482, 4303, \dodoi{10.1093/mnras/sty2936}

\bibitem[{{McKinney} \& {Gammie}(2004)}]{mckinney2004}
{McKinney}, J.~C., \& {Gammie}, C.~F. 2004, \apj, 611, 977,
  \dodoi{10.1086/422244}

\bibitem[{{Meisenheimer} {et~al.}(1997){Meisenheimer}, {Yates}, \&
  {Roeser}}]{meisenheimer1997}
{Meisenheimer}, K., {Yates}, M.~G., \& {Roeser}, H.~J. 1997, \aap, 325, 57

\bibitem[{{Meliani} \& {Keppens}(2009)}]{meliani2009}
{Meliani}, Z., \& {Keppens}, R. 2009, \apj, 705, 1594,
  \dodoi{10.1088/0004-637X/705/2/1594}

\bibitem[{{Mignone} {et~al.}(2007){Mignone}, {Bodo}, {Massaglia}, {Matsakos},
  {Tesileanu}, {Zanni}, \& {Ferrari}}]{mignone2007pluto}
{Mignone}, A., {Bodo}, G., {Massaglia}, S., {et~al.} 2007, \apjs, 170, 228,
  \dodoi{10.1086/513316}

\bibitem[{{Mignone} \& {McKinney}(2007)}]{mignone2007taub}
{Mignone}, A., \& {McKinney}, J.~C. 2007, \mnras, 378, 1118,
  \dodoi{10.1111/j.1365-2966.2007.11849.x}

\bibitem[{{Mignone} {et~al.}(2005){Mignone}, {Plewa}, \& {Bodo}}]{Mignone2005}
{Mignone}, A., {Plewa}, T., \& {Bodo}, G. 2005, \apjs, 160, 199,
  \dodoi{10.1086/430905}

\bibitem[{{Mignone} {et~al.}(2010){Mignone}, {Rossi}, {Bodo}, {Ferrari}, \&
  {Massaglia}}]{Mignone2010}
{Mignone}, A., {Rossi}, P., {Bodo}, G., {Ferrari}, A., \& {Massaglia}, S. 2010,
  \mnras, 402, 7, \dodoi{10.1111/j.1365-2966.2009.15642.x}

\bibitem[{{Mizuno} {et~al.}(2007){Mizuno}, {Hardee}, \&
  {Nishikawa}}]{mizuno2007}
{Mizuno}, Y., {Hardee}, P., \& {Nishikawa}, K.-I. 2007, \apj, 662, 835,
  \dodoi{10.1086/518106}

\bibitem[{{Mizuno} {et~al.}(2014){Mizuno}, {Pohl}, {Niemiec}, {Zhang},
  {Nishikawa}, \& {Hardee}}]{mizuno2014}
{Mizuno}, Y., {Pohl}, M., {Niemiec}, J., {et~al.} 2014, \mnras, 439, 3490,
  \dodoi{10.1093/mnras/stu196}

\bibitem[{{Mukherjee} {et~al.}(2021){Mukherjee}, {Bodo}, {Rossi}, {Mignone}, \&
  {Vaidya}}]{Mukherjee2021}
{Mukherjee}, D., {Bodo}, G., {Rossi}, P., {Mignone}, A., \& {Vaidya}, B. 2021,
  \mnras, 505, 2267, \dodoi{10.1093/mnras/stab1327}

\bibitem[{{Ortu{\~n}o-Mac{\'\i}as} {et~al.}(2022){Ortu{\~n}o-Mac{\'\i}as},
  {Nalewajko}, {Uzdensky}, {Begelman}, {Werner}, {Chen}, \&
  {Mishra}}]{ortuno-macias2022}
{Ortu{\~n}o-Mac{\'\i}as}, J., {Nalewajko}, K., {Uzdensky}, D.~A., {et~al.}
  2022, \apj, 931, 137, \dodoi{10.3847/1538-4357/ac6acd}

\bibitem[{{Pierre Auger Collaboration} {et~al.}(2017){Pierre Auger
  Collaboration}, {Aab}, {Abreu}, {Aglietta}, {Samarai}, {Albuquerque},
  {Allekotte}, {Almela}, {Alvarez Castillo}, {Alvarez-Mu{\~n}iz}, {Anastasi},
  {Anchordoqui}, {Andrada}, {Andringa}, {Aramo}, {Arqueros}, {Arsene},
  {Asorey}, {Assis}, {Aublin}, {Avila}, {Badescu}, {Balaceanu}, {Barbato},
  {Barreira Luz}, {Beatty}, {Becker}, {Bellido}, {Berat}, {Bertaina}, {Bertou},
  {Biermann}, {Billoir}, {Biteau}, {Blaess}, {Blanco}, {Blazek}, {Bleve},
  {Boh{\'a}{\v{c}}ov{\'a}}, {Boncioli}, {Bonifazi}, {Borodai}, {Botti},
  {Brack}, {Brancus}, {Bretz}, {Bridgeman}, {Briechle}, {Buchholz}, {Bueno},
  {Buitink}, {Buscemi}, {Caballero-Mora}, {Caccianiga}, {Cancio}, {Canfora},
  {Caramete}, {Caruso}, {Castellina}, {Cataldi}, {Cazon}, {Chavez},
  {Chinellato}, {Chudoba}, {Clay}, {Cobos}, {Colalillo}, {Coleman}, {Collica},
  {Coluccia}, {Concei{\c{c}}{\~a}o}, {Consolati}, {Contreras}, {Cooper},
  {Coutu}, {Covault}, {Cronin}, {D'Amico}, {Daniel}, {Dasso}, {Daumiller},
  {Dawson}, {de Almeida}, {de Jong}, {De Mauro}, {de Mello Neto}, {De Mitri},
  {de Oliveira}, {de Souza}, {Debatin}, {Deligny}, {Di Giulio}, {Di Matteo},
  {D{\'\i}az Castro}, {Diogo}, {Dobrigkeit}, {D'Olivo}, {Dorosti}, {dos Anjos},
  {Dova}, {Dundovic}, {Ebr}, {Engel}, {Erdmann}, {Erfani}, {Escobar},
  {Espadanal}, {Etchegoyen}, {Falcke}, {Farrar}, {Fauth}, {Fazzini}, {Fenu},
  {Fick}, {Figueira}, {Filip{\v{c}}i{\v{c}}}, {Fratu}, {Freire}, {Fujii},
  {Fuster}, {Gaior}, {Garc{\'\i}a}, {Garcia-Pinto}, {Gat{\'e}}, {Gemmeke},
  {Gherghel-Lascu}, {Ghia}, {Giaccari}, {Giammarchi}, {Giller}, {G{\l}as},
  {Glaser}, {Golup}, {G{\'o}mez Berisso}, {G{\'o}mez Vitale}, {Gonz{\'a}lez},
  {Gorgi}, {Gorham}, {Grillo}, {Grubb}, {Guarino}, {Guedes}, {Hampel},
  {Hansen}, {Harari}, {Harrison}, {Harton}, {Haungs}, {Hebbeker}, {Heck},
  {Heimann}, {Herve}, {Hill}, {Hojvat}, {Holt}, {Homola}, {H{\"o}randel},
  {Horvath}, {Hrabovsk{\'y}}, {Huege}, {Hulsman}, {Insolia}, {Isar}, {Jandt},
  {Jansen}, {Johnsen}, {Josebachuili}, {Jurysek}, {K{\"a}{\"a}p{\"a}},
  {Kambeitz}, {Kampert}, {Katkov}, {Keilhauer}, {Kemmerich}, {Kemp}, {Kemp},
  {Kieckhafer}, {Klages}, {Kleifges}, {Kleinfeller}, {Krause}, {Krohm},
  {Kuempel}, {Kukec Mezek}, {Kunka}, {Kuotb Awad}, {LaHurd}, {Lauscher},
  {Legumina}, {Leigui de Oliveira}, {Letessier-Selvon}, {Lhenry-Yvon}, {Link},
  {Lo Presti}, {Lopes}, {L{\'o}pez}, {L{\'o}pez Casado}, {Luce}, {Lucero},
  {Malacari}, {Mallamaci}, {Mandat}, {Mantsch}, {Mariazzi}, {Mari{\c{s}}},
  {Marsella}, {Martello}, {Martinez}, {Mart{\'\i}nez Bravo}, {Mas{\'\i}as
  Meza}, {Mathes}, {Mathys}, {Matthews}, {Matthews}, {Matthiae}, {Mayotte},
  {Mazur}, {Medina}, {Medina-Tanco}, {Melo}, {Menshikov}, {Merenda}, {Michal},
  {Micheletti}, {Middendorf}, {Miramonti}, {Mitrica}, {Mockler}, {Mollerach},
  {Montanet}, {Morello}, {Mostaf{\'a}}, {M{\"u}ller}, {M{\"u}ller}, {Muller},
  {M{\"u}ller}, {Mussa}, {Naranjo}, {Nellen}, {Nguyen}, {Niculescu-Oglinzanu},
  {Niechciol}, {Niemietz}, {Niggemann}, {Nitz}, {Nosek}, {Novotny},
  {No{\v{z}}ka}, {N{\'u}{\~n}ez}, {Ochilo}, {Oikonomou}, {Olinto}, {Palatka},
  {Pallotta}, {Papenbreer}, {Parente}, {Parra}, {Paul}, {Pech}, {Pedreira},
  {Pkala}, {Pelayo}, {Pe{\~n}a-Rodriguez}, {Pereira}, {Perl{\'\i}n}, {Perrone},
  {Peters}, {Petrera}, {Phuntsok}, {Piegaia}, {Pierog}, {Pieroni}, {Pimenta},
  {Pirronello}, {Platino}, {Plum}, {Porowski}, {Prado}, {Privitera}, {Prouza},
  {Quel}, {Querchfeld}, {Quinn}, {Ramos-Pollan}, {Rautenberg}, {Ravignani},
  {Revenu}, {Ridky}, {Riehn}, {Risse}, {Ristori}, {Rizi}, {Rodrigues de
  Carvalho}, {Rodriguez Fernandez}, {Rodriguez Rojo}, {Rogozin}, {Roncoroni},
  {Roth}, {Roulet}, {Rovero}, {Ruehl}, {Saffi}, {Saftoiu}, {Salamida},
  {Salazar}, {Saleh}, {Salesa Greus}, {Salina}, {S{\'a}nchez}, {Sanchez-Lucas},
  {Santos}, {Santos}, {Sarazin}, {Sarmento}, {Sarmiento}, {Sato}, {Schauer},
  {Scherini}, {Schieler}, {Schimp}, {Schmidt}, {Scholten}, {Schov{\'a}nek},
  {Schr{\"o}der}, {Schulz}, {Schumacher}, {Sciutto}, {Segreto}, {Settimo},
  {Shadkam}, {Shellard}, {Sigl}, {Silli}, {Sima}, {{\'S}mia{\l}kowski},
  {{\v{S}}m{\'\i}da}, {Snow}, {Sommers}, {Sonntag}, {Sorokin}, {Squartini},
  {Stanca}, {Stani{\v{c}}}, {Stasielak}, {Stassi}, {Strafella}, {Suarez},
  {Suarez Dur{\'a}n}, {Sudholz}, {Suomij{\"a}rvi}, {Supanitsky},
  {{\v{S}}up{\'\i}k}, {Swain}, {Szadkowski}, {Taboada}, {Taborda}, {Tapia},
  {Theodoro}, {Timmermans}, {Todero Peixoto}, {Tomankova}, {Tom{\'e}},
  {Torralba Elipe}, {Travnicek}, {Trini}, {Ulrich}, {Unger}, {Urban},
  {Vald{\'e}s Galicia}, {Vali{\~n}o}, {Valore}, {van Aar}, {van Bodegom}, {van
  den Berg}, {van Vliet}, {Varela}, {Vargas C{\'a}rdenas}, {Varner},
  {V{\'a}zquez}, {Veberi{\v{c}}}, {Ventura}, {Vergara Quispe}, {Verzi},
  {Vicha}, {Villase{\~n}or}, {Vorobiov}, {Wahlberg}, {Wainberg}, {Walz},
  {Watson}, {Weber}, {Weindl}, {Wiencke}, {Wilczy{\'n}ski}, {Wirtz},
  {Wittkowski}, {Wundheiler}, {Yang}, {Yushkov}, {Zas}, {Zavrtanik},
  {Zavrtanik}, {Zepeda}, {Zimmermann}, {Ziolkowski}, {Zong}, \&
  {Zuccarello}}]{PAC2017}
{Pierre Auger Collaboration}, {Aab}, A., {Abreu}, P., {et~al.} 2017, Science,
  357, 1266, \dodoi{10.1126/science.aan4338}

\bibitem[{{Porth} \& {Fendt}(2010)}]{porth2010}
{Porth}, O., \& {Fendt}, C. 2010, \apj, 709, 1100,
  \dodoi{10.1088/0004-637X/709/2/1100}

\bibitem[{{Qian} {et~al.}(2018){Qian}, {Fendt}, \& {Vourellis}}]{qian2018}
{Qian}, Q., {Fendt}, C., \& {Vourellis}, C. 2018, \apj, 859, 28,
  \dodoi{10.3847/1538-4357/aabd36}

\bibitem[{{Reville} \& {Bell}(2014)}]{reville2014}
{Reville}, B., \& {Bell}, A.~R. 2014, \mnras, 439, 2050,
  \dodoi{10.1093/mnras/stu088}

\bibitem[{{Rieger} {et~al.}(2007){Rieger}, {Bosch-Ramon}, \&
  {Duffy}}]{rieger2007}
{Rieger}, F.~M., {Bosch-Ramon}, V., \& {Duffy}, P. 2007, \apss, 309, 119,
  \dodoi{10.1007/s10509-007-9466-z}

\bibitem[{{Rossi} {et~al.}(2008){Rossi}, {Mignone}, {Bodo}, {Massaglia}, \&
  {Ferrari}}]{rossi2008}
{Rossi}, P., {Mignone}, A., {Bodo}, G., {Massaglia}, S., \& {Ferrari}, A. 2008,
  \aap, 488, 795, \dodoi{10.1051/0004-6361:200809687}

\bibitem[{{Sheikhnezami} \& {Fendt}(2015)}]{sheikhnezami2015}
{Sheikhnezami}, S., \& {Fendt}, C. 2015, \apj, 814, 113,
  \dodoi{10.1088/0004-637X/814/2/113}

\bibitem[{{Sheikhnezami} {et~al.}(2012){Sheikhnezami}, {Fendt}, {Porth},
  {Vaidya}, \& {Ghanbari}}]{sheikhnezami2012}
{Sheikhnezami}, S., {Fendt}, C., {Porth}, O., {Vaidya}, B., \& {Ghanbari}, J.
  2012, \apj, 757, 65, \dodoi{10.1088/0004-637X/757/1/65}

\bibitem[{{Sironi} {et~al.}(2021){Sironi}, {Rowan}, \& {Narayan}}]{sironi2021}
{Sironi}, L., {Rowan}, M.~E., \& {Narayan}, R. 2021, \apjl, 907, L44,
  \dodoi{10.3847/2041-8213/abd9bc}

\bibitem[{{Stepanovs} \& {Fendt}(2016)}]{stepanovs2016}
{Stepanovs}, D., \& {Fendt}, C. 2016, \apj, 825, 14,
  \dodoi{10.3847/0004-637X/825/1/14}

\bibitem[{{Summerlin} \& {Baring}(2012)}]{summerlin2012}
{Summerlin}, E.~J., \& {Baring}, M.~G. 2012, \apj, 745, 63,
  \dodoi{10.1088/0004-637X/745/1/63}

\bibitem[{Synge(1957)}]{Synge1957}
Synge, J.~L. 1957, The relativistic gas (North-Holland Publishing Company,
  Amsterdam; Interscience Publishers Inc., New York), xi+108

\bibitem[{{Tchekhovskoy} {et~al.}(2011){Tchekhovskoy}, {Narayan}, \&
  {McKinney}}]{tchekhovskoy2011}
{Tchekhovskoy}, A., {Narayan}, R., \& {McKinney}, J.~C. 2011, \mnras, 418, L79,
  \dodoi{10.1111/j.1745-3933.2011.01147.x}

\bibitem[{{Vaidya} {et~al.}(2018){Vaidya}, {Mignone}, {Bodo}, {Rossi}, \&
  {Massaglia}}]{vaidya2018}
{Vaidya}, B., {Mignone}, A., {Bodo}, G., {Rossi}, P., \& {Massaglia}, S. 2018,
  \apj, 865, 144, \dodoi{10.3847/1538-4357/aadd17}

\bibitem[{{von Fellenberg} {et~al.}(2023){von Fellenberg}, {Janssen},
  {Davelaar}, {Zaja{\v{c}}ek}, {Britzen}, {Falcke}, {K{\"o}rding}, \&
  {Ros}}]{vonFellenberg2023}
{von Fellenberg}, S.~D., {Janssen}, M., {Davelaar}, J., {et~al.} 2023, arXiv
  e-prints, arXiv:2303.00603, \dodoi{10.48550/arXiv.2303.00603}

\bibitem[{{Vourellis} \& {Fendt}(2021)}]{vourellis2021}
{Vourellis}, C., \& {Fendt}, C. 2021, \apj, 911, 85,
  \dodoi{10.3847/1538-4357/abe93b}

\bibitem[{{Vourellis} {et~al.}(2019){Vourellis}, {Fendt}, {Qian}, \&
  {Noble}}]{vourellis2019}
{Vourellis}, C., {Fendt}, C., {Qian}, Q., \& {Noble}, S.~C. 2019, \apj, 882, 2,
  \dodoi{10.3847/1538-4357/ab32e2}

\bibitem[{{Webb}(1989)}]{webb1989}
{Webb}, G.~M. 1989, \apj, 340, 1112, \dodoi{10.1086/167462}

\bibitem[{{Yates} {et~al.}(2018){Yates}, {Shabala}, \& {Krause}}]{yates2018}
{Yates}, P.~M., {Shabala}, S.~S., \& {Krause}, M. G.~H. 2018, \mnras, 480,
  5286, \dodoi{10.1093/mnras/sty2191}

\bibitem[{{Zanni} {et~al.}(2007){Zanni}, {Ferrari}, {Rosner}, {Bodo}, \&
  {Massaglia}}]{zanni2007}
{Zanni}, C., {Ferrari}, A., {Rosner}, R., {Bodo}, G., \& {Massaglia}, S. 2007,
  \aap, 469, 811, \dodoi{10.1051/0004-6361:20066400}

\end{thebibliography}
\bibliographystyle{aasjournal}

\appendix
\counterwithin{figure}{section}

\section{Jet Injection Profiles}
\label{appendix: inj_profiles}
In our study, we adopt the equilibrium solution derived by \citet{bodo2019}, 
who solved for the radial component of the momentum equation for constant gas pressure $p$.

In cylindrical coordinates ($r$, $\phi$, $z$) they find
\begin{equation}
    \rho \gamma^2 v_{\phi}^2 = \frac{1}{2r}\frac{d(r^2H^2)}{dr}+\frac{r}{2}\frac{dB_z^2}{dr},
    \nonumber
\end{equation}
where $H^2 = B_{\phi}^2 - E_r^2$ and $E_r = v_z B_{\phi}-v_{\phi}B_z$.
One may define a pitch angle of the magnetic field, $\delta = r B_z / B_{\phi}$,
the Lorentz factor in $z-$direction,
\begin{equation}
\label{eq:gamma_prof}
    \gamma_z (r) = 1 + \frac{\gamma_c -1}{\cosh{(r/r_j)}^6},
    \nonumber
\end{equation}
the azimuthal velocity,
\begin{equation}
    \gamma^2v_{\phi}^2 = r^2 \Omega_c^2 \gamma_c^2 \exp\left(-\frac{r^4}{a^4}\right),
    \nonumber
\end{equation}
and the function 
\begin{equation}
    H^2 = \frac{H_c^2}{r^2}\left[1 - \exp\left(-\frac{r^4}{a^4}\right)\right].
    \nonumber
\end{equation}
Here, 
$r$ is the radial coordinate in the cylindrical coordinate system, 
$\gamma_c$ is the Lorentz factor at $r=0$, 
$r_j = 1.0$ is the radius of the injection nozzle, 
$a = 0.6$ is {\rev commonly denoted as the {\em magnetization radius}} at which the magnetization of the jet nozzle has its maximum value,
and $\Omega_c$ is the angular velocity.  
{\rev Choosing a magnetization radius $a$ less than the jet radius $r_{\rm j}$, 
we make sure that we do not have an infinite Lorentz force or a strong shear acting at the jet edges.}
Note that a subscript $c$ denotes the value of a parameter at $r = 0$.

\begin{figure*}[t]
    \centering
    \includegraphics[width = \linewidth]{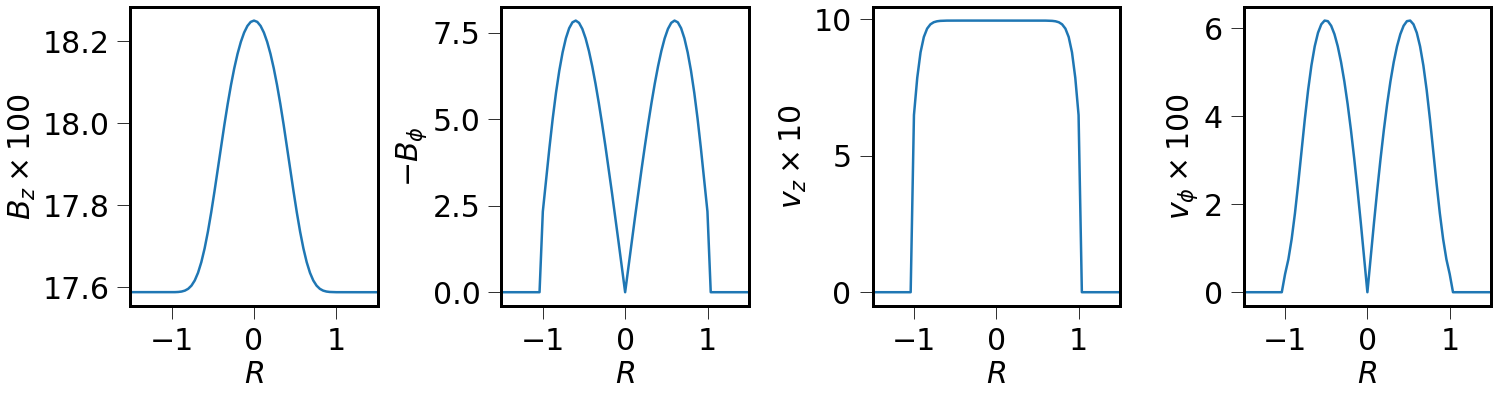}
    \caption{Variation of the $z-$ and $\phi-$components of the velocity and magnetic field in the steady injection nozzle as a function of radius $R$.}
    \label{fig:profiles}
\end{figure*}

This gives the following injection profiles within the cylindrical injection nozzle (see \citealt{bodo2019} for detailed derivation)
\begin{equation}
    v_{\phi}^2(r) = \frac{r^2\gamma_c^2\Omega_c^2}{\gamma_z^2} \left[1 + r^2\gamma_c^2 \Omega_c^2 \exp\left(-\frac{r^4}{a^4}\right) \right]^{-1} \exp\left(-\frac{r^4}{a^4}\right),
    \nonumber
\end{equation}

\begin{equation}
    v_r = B_r = 0,
    \nonumber
\end{equation}

\begin{equation}
    B_z^2 = B_{zc}^2 - (1-\zeta) \frac{H_c^2 \sqrt{\pi}}{a^2} \mathcal{E}\left(\frac{r^2}{a^2}\right),
    \nonumber
\end{equation}

\begin{equation}
\label{eq:b_phi}
    B_{\phi} (r) = \frac{-v_{\phi} v_z B_z - \sqrt{v_{\phi}B_z^2 + H^2(1-v_z^2)}}{1-v_z^2},
\end{equation}
where
\begin{equation}
    a^4 B_{zc}^2 = \frac{H_c^2 \delta_c^2}{1 - (\delta_c\Omega_c + v_{zc})^2},\quad\quad   \zeta \equiv \frac{\rho_{\rm j} \gamma_c^2 \Omega^2 a^4}{2 H_c^2}.\nonumber
\end{equation}
The parameter $\zeta$ governs the strength of rotation and $\mathcal{E}$ is the error function. 
Note that not all combinations of input parameters lead to a physical solution. 
In general, it must be required that $B_z^2$ and $B_{\phi}^2$ are positive everywhere. 

Overall, with these equations we can define a rotating, magnetized, relativistic jet of density $\rho_{\rm j}$ and pressure $p$, 
launched from the injection nozzle, which is propagating along the $z-$direction and is governed by the choice
of the parameters 
$\rho_{\rm j}$, $\gamma_c$, $\Omega_c$, $B_{zc}$, $\delta_c$ and $p$. 
Our choice of these parameters for our simulations is summarized in Table~\ref{tbl:input}.
We show the profiles of the $z-$ and $\phi-$components of the velocity and magnetic field injected through the steady jet nozzle setup as a function of radius $R$, following the aforementioned equations in Fig.~\ref{fig:profiles}.

\section{Resolution Study}
\label{appendix:resolution}
A resolution study is essential for numerical simulation work.
It is important for our study in particular as we are not only interested in the gross dynamical behavior of the jet flow,
but also in its small-scale structure that is eventually responsible for particle acceleration.
Thus, it is necessary to avoid (artificial) numerical effects in our study and achieve a sufficient resolution of turbulent structures.

\begin{figure*}[t]
    \centering
    \includegraphics[width = 0.233\linewidth]{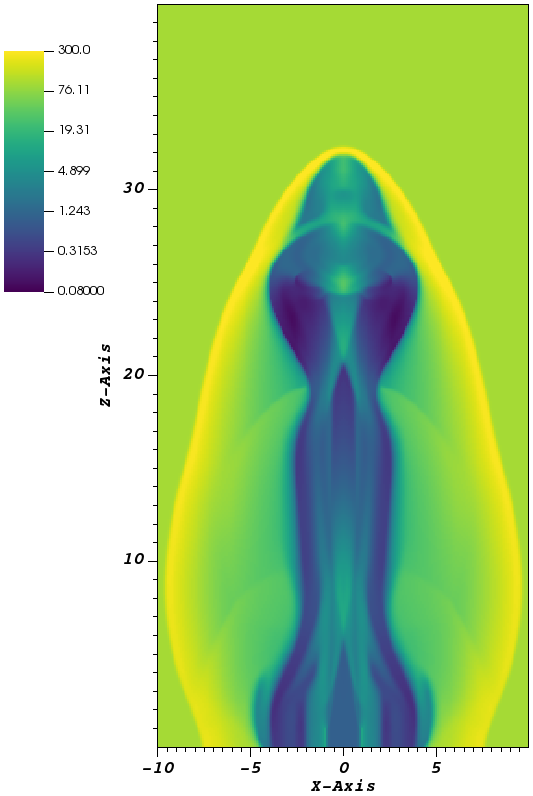}
    \includegraphics[width = 0.18\linewidth]{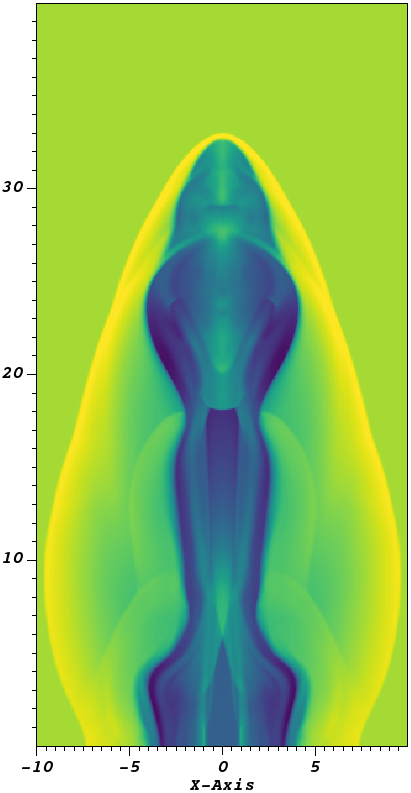}
    \includegraphics[width = 0.18\linewidth]{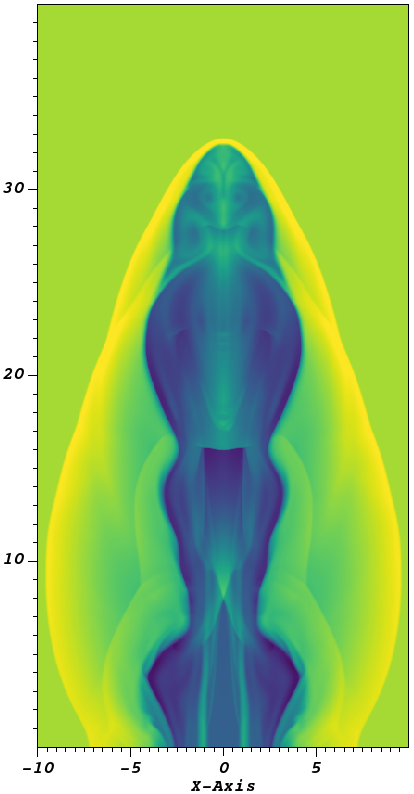}
    \includegraphics[width = 0.18\linewidth]{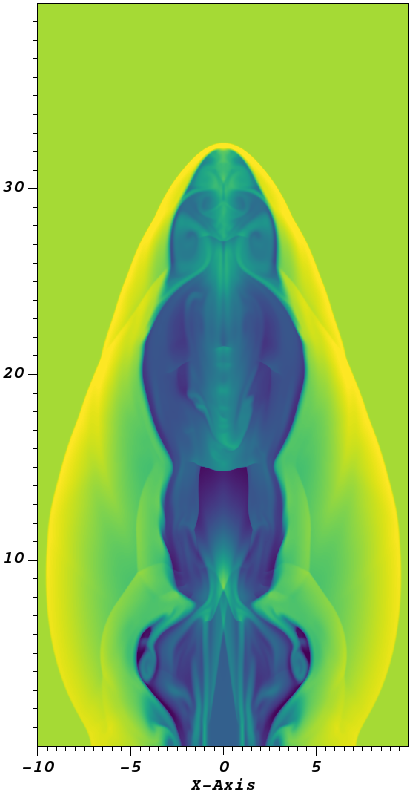}
    \includegraphics[width = 0.18\linewidth]{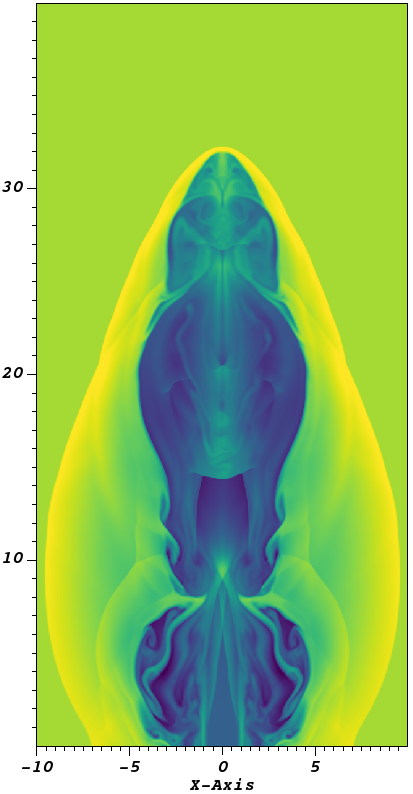}
    \caption{Distribution of density (in $\log$ scale) in the domain at a time $t = 50$ (in code units) in the $x-z$ plane (for $y=0$) for the simulation run {\em std} at a resolution of (from left to right) 10, 15, 20, 25 and 30 cells per jet radius.}
    \label{fig:res_rho}
\end{figure*}

We have performed a number of simulation runs with different numerical resolution applying the setup with the steady jet nozzle.
To this end, defining the resolution as the number of grid cells inside the jet radius $r_j$,
we run simulations with a resolution of $10$, $15$, $20$, $25$ and $30$ pixels per $r_j$. 
In Fig~\ref{fig:res_rho} we show 2D slices of the density distribution in the $x-z$ plane (at $y=0$) at time $t=50$
for various resolutions for the simulation setup \textit{std}, in order to compare the effect of resolution on the dynamics of the jet. 

We see that with a low resolution of 10 grid cells per jet radius, we miss a lot of small-scale structure of the jet which could be essential for particle acceleration. 
Also, we do not see the formation of the strong steady {\rev internal} shock structure, which is a very efficient site of particle acceleration as we discussed. 
The formation of the strong steady {\rev internal} shock can be detected as we move to a higher resolution of 15 grid cells per jet radius. 
Still the position of the strong, steady {\rev internal} shock still depends on resolution, which converges only even higher resolution. 
As we go on from a resolution of 15 to a resolution of 20 cells per $r_j$, we see additional shocks near the the jet head seen as an over-dense region of filamentary structure. 
In addition we see an upstream shifting of the strong, steady {\rev internal} shock. 

These differences to the lower resolution study are even more visible as we move to a resolution of 25 cells per $r_j$. 
Here, we see the formation of two finger-like (crab-like?) structures near the base of the jet, which is not present at lower resolutions. 
We also see the location of the {\rev strong steady  internal shock} further upstream, although now the difference in position is smaller than between the lower resolutions. 
We thus think that we have rule out simulation runs applying a resolution of less than 25 cells per $r_j$.
On the other hand, runs with a resolution of 25 and 30 cells per $r_j$ look really similar, suggesting a convergence with regards to the resolution.

\begin{figure}[t]
    \centering
    \includegraphics[width = \linewidth]{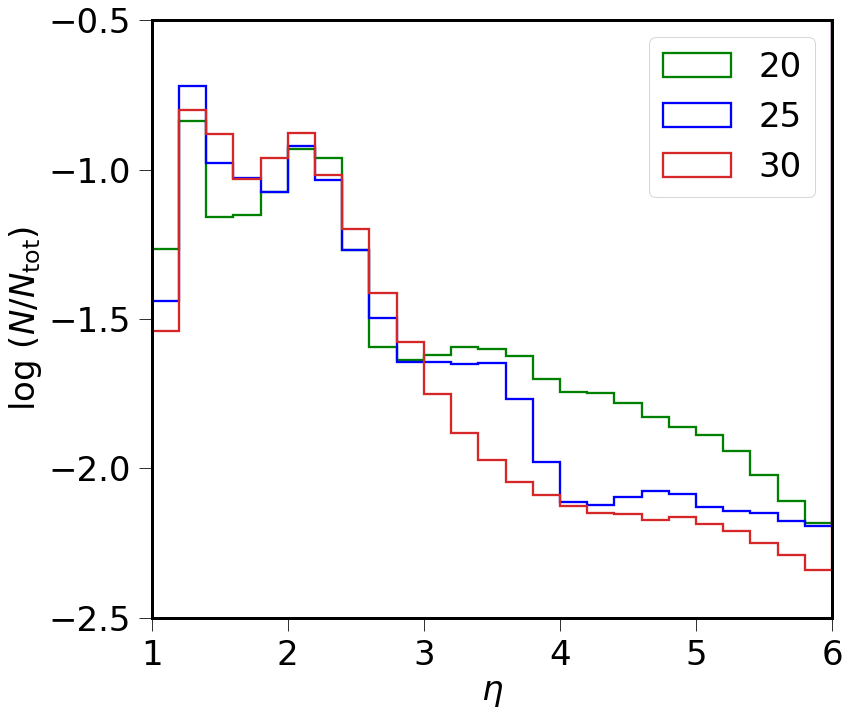}
    \caption{Histogram of compression ratios $\eta$. 
    Shown is the number of particles $N$  in a specific range of $\eta$, normalized by the total number of particles in the domain $N_{\rm tot}$
    (in $\log$ scale) for different compression ratio for the simulation run \textit{std} at a resolution of 
    20 (\textit{green}), 
    25 (\textit{blue}) and 
    30 (\textit{red}) at time $t =50$. 
    }
    \label{fig:res_cmpr}
\end{figure}

We also investigated convergence beyond the dynamical jet structure.
In particular, we are interested in studying how the particle acceleration is affected, for example by effects that remain invisible by 
a comparison of the pure dynamical structure. 
In Fig.~\ref{fig:res_cmpr} we show the histograms of the shock compression ratio $\eta$ the particles have experienced until $t=50$ for the simulation setup \textit{std} 
for the different resolutions of 20, 25, and 30 grid cells per jet radius .
We clearly see that the difference in the statistics between the runs with resolution 25 and 30 grid cells per jet radius is lower than that between the runs 
with resolution of 20 and 25 grid cells per jet radius.

We conclude that for both, the dynamical evolution and the particle evolution, the simulations with resolution 25 grid cells per jet radius are sufficiently
converged compared to the resolution of 30 grid cells per jet radius, which justifies the use of a resolution of 25 grid cells per jet radius.

\end{document}